\documentclass[iop,apj,numberedappendix,twocolappendix]{emulateapj}

\usepackage{natbib}
\usepackage{amsmath,empheq}
\usepackage{multirow,booktabs,threeparttablex}
\usepackage{color}
\usepackage{hyperref}
\hypersetup{
  pdfinfo={
    Title={A New Optical Polarization Catalog for the Small Magellanic Cloud: The Magnetic Field Structure},
    Author={A. Lobo Gomes, A. M. Magalhaes, A. Pereyra, C. V. Rodrigues},
    Subject={Astrophysics},
    Keywords={galaxies: ISM, Magellanic Clouds, magnetic fields --- techniques: polarimetric}
  }
}

\newcommand{\noprint}[1]{}
\newcommand{\figsetstart}{}
\newcommand{\figsetend}{}
\newcommand{\figsetgrpstart}{}
\newcommand{\figsetgrpend}{}

\newcommand{\figsetgrpnum}[1]{\noprint{#1}}

\newcommand{\figsetplot}[1]{\noprint{#1}}
\newcommand{\figsetgrpnote}[1]{\noprint{#1}}

\shorttitle{A New Optical Polarization Catalog for the SMC: The Magnetic Field Structure}
\shortauthors{Lobo Gomes, Magalh\~aes, Pereyra \& Rodrigues}
\slugcomment{Submitted: December 08, 2014; Accepted: April 7, 2015}

\begin{document}

\title{A New Optical Polarization Catalog for the Small Magellanic Cloud:\\ The Magnetic Field Structure}

\author{Aiara Lobo Gomes$^{1,2,3}$, Ant\^onio M\'ario Magalh\~aes$^{3}$, Antonio Pereyra$^{4}$ \& Cl\'audia Vilega Rodrigues$^{5}$} 

\affil{$^{1}$Max-Planck-Institut f\"ur Astronomie, K\"onigstuhl 17, D-69117, Heidelberg, Germany}
\affil{$^{2}$Member of the International Max Planck Research School for Astronomy and Cosmic Physics at the University of Heidelberg}
\affil{$^{3}$Instituto de Astronomia, Geof\'\i sica e Ci\^encias Atmosf\'ericas, Universidade de S\~ao Paulo, Rua do Mat\~ao 1226, 05508-900, S\~ao Paulo, SP, Brazil}
\affil{$^{4}$Instituto Geof\'isico del Per\'u, \'Area Astronom\'ia, Calle Badajoz, Lima 3, Lima, Per\'u}
\affil{$^{5}$Instituto Nacional de Pesquisas Espaciais, Avenida dos Astronautas 1758, 12227-010, S\~ao Jos\'e dos Campos, SP, Brazil}

\email[email: ]{gomes@mpia.de}


\begin{abstract}

We present a new optical polarimetric catalog for the Small Magellanic Cloud (SMC). It contains a total of 7207 stars, located in the northeast (NE) and
Wing sections of the SMC and part of the Magellanic Bridge. This new catalog is a significant improvement compared to previous polarimetric catalogs for
the SMC. We used it to study the
sky-projected interstellar magnetic field structure of the SMC. Three trends were observed for the ordered magnetic field direction at
position angles (PAs) of ($65^{\circ}\pm10^{\circ}$), ($115^{\circ}\pm10^{\circ}$), and ($150^{\circ}\pm10^{\circ}$). Our results suggest the existence of an ordered magnetic field aligned with the Magellanic Bridge
direction and SMC's Bar in the NE region, which have PAs roughly at $115^{\circ}.4$ and $45^{\circ}$, respectively. However, the
overall magnetic field
structure is fairly complex. The trends at $115^{\circ}$ and $150^{\circ}$ may be correlated with the SMC's bimodal
structure, observed in Cepheids' distances and HI velocities. We derived a value of $B_{\text{sky}}~=~(0.947\pm0.079)~\mu$G for the ordered
sky-projected magnetic field, and $\delta B~=~(1.465\pm0.069)~\mu$G for
the turbulent magnetic field. This estimate of $B_{\text{sky}}$ is significantly larger (by a factor of $\sim 10$) than the line of sight field derived from Faraday
rotation observations, suggesting that most of the ordered field component is on the plane of the sky. A turbulent magnetic field stronger than the ordered field agrees with observed
estimates for other irregular and spiral galaxies. For the SMC the $B_{sky}/\delta B$ ratio is closer to what is observed for our
Galaxy than other irregular dwarf galaxies.

\end{abstract}



\section{Introduction}\label{sec:intro}

	The Small Magellanic Cloud (SMC) and the Large Magellanic Cloud (LMC) are gas rich irregular galaxies and satellites of the Milky Way (MW). Given
the proximity to the Galaxy and their high Galactic
latitude, the Magellanic Clouds (MCs) are perfect laboratories for extragalactic studies, since the light emitted from them is not much attenuated
by the dust present in the Galactic disk. These objects are highly studied, motivated by the fact that, through understanding them, we can also better comprehend, for instance, galaxy evolution
from high redshifts to now; the structure of irregular dwarf galaxies; the interstellar medium (ISM) content at the external parts of huge spiral galaxies; star formation at low
metalicity environments; dwarf galaxies in interaction; and the relation of satellite galaxies with their hosts.

	The MCs are low metalicity galaxies -- $0.5Z_{\odot}$ and $0.2Z_{\odot}$ for the LMC and SMC, respectively \citep{bib:kurt98}. This peculiarity makes their
ISM particularly different from the one of our Galaxy. Among the major differences, we can point out the high gas-to-dust ratio
\citep{bib:stanimirovic00,bib:bot04,bib:meixner10} and the submm excess emission \citep{bib:bernard08,bib:israel10,bib:bot10,bib:planck11q}.

	The SMC consists of two main features: Bar and Wing. The Bar component is the main body of the SMC and corresponds to its
northern part, which
has a position angle (PA) of $\sim45^{\circ}$. The Wind is located $\sim2^{\circ}$ East of the Bar and corresponds to its southern part \citep{bib:bergh07}.

	The dynamical evolution of the SMC--LMC--MW system seems to have generated two interesting structures -- the Magellanic Stream and the
Magellanic Bridge. These structures are well illustrated in the HI column density map (Figure 2) presented by \cite{bib:bruns05}. The Magellanic Stream is a
gas tail covering over $100^{\circ}$, also going through the Galactic south pole. One scenario says that this gas might have been pulled out of the
SMC and LMC due to tidal interaction between the MCs and the MW \citep{bib:putman03}. On the other side, the Magellanic Bridge seems to have been generated due to the same effect, but between the SMC
and the LMC \citep{bib:bekki09}. More recently, high precision proper motions from the {\it Hubble Space Telescope} showed that the MCs are moving
$80$~km/s faster than earlier estimates \citep{bib:kallivayalil06a,bib:kallivayalil06b}. These high velocities favor a scenario where the MCs are in their first infall toward the
MW \citep{bib:besla07,bib:boylan11,bib:busha11}. In this scenario, \cite{bib:besla12} found that the Magellanic Stream may have been formed by LMC tides on the SMC before the system was accreted by the MW.
\cite{bib:diaz12} studied several orbits consistent with the recent proper motions. They assumed an orbital history where the MCs became a strongly
interacting pair just recently. In this picture, their first close encounter provided enough tidal forces to disrupt the SMC's disk and create the
Magellanic Stream. In contrast to the Magellanic Stream, the Magellanic Bridge has
a young stellar population \citep{bib:irwin85}, besides being composed of gas and dust
\citep{bib:hindman63,bib:sofue94,bib:muller04,bib:gordon09}. The study of gas-to-dust ratio
of the Magellanic Bridge confirms the hypothesis that the Magellanic Bridge is formed of
material that was stripped from the SMC \citep{bib:gordon09}. Moreover it was found that at the Magellanic Bridge the gas-to-dust ratio is about
$12$ times higher than in the Galaxy and about two times higher than the value expected by its metalicity, indicating that some of the dust in that region must have been
destroyed.

	It is not well known whether the MCs were formed as a binary system or were dynamically coupled about $3$~Gyr ago \citep{bib:bekki05}. The kinematic
history reconstruction of the MCs suggests that their last closest approach occurred about $0.2$~Gyr ago, when they came within 2--7~kpc distance of each other, a likely progenitor event for the
construction of the Bridge. Another possible result of this interaction is the stretching in northeastern and Wing sections toward the LMC \citep{bib:westerlund91}.

	The last interaction was also likely responsible for the bimodality of the SMC along its depth \citep{bib:mf84}, a feature that is corroborated by the velocity components in HI data, which are
$40$~km s$^{-1}$ in difference \citep{bib:mathewson84,bib:stanimirovic04}. Distances to Cepheids show that the high velocity component is located
further in distance, while the low velocity component is closer. Both components appear to be expanding at around $15$~km~s$^{-1}$. The low velocity component itself is further subdivided into two
other groups, one extending from the center to the northeast (NE) at a location of $50$~kpc, the other, extending from the Bar to the southwest (SW) at a location of $60$~kpc. The Bar is the most
extended region ($45$--$90$~kpc) and therefore the HI profiles in that direction are very complex \citep{bib:mf84}. The study of the HI component at
the SMC shows that it rotates
\citep{bib:stanimirovic04}, indicating that the SMC contains a disk component.

	Distances to Cepheids also show that the two gas components are about $10$~kpc apart and have a
depth of $17$~kpc. A double component, relating to the distances, is also observed in Cepheids located in the Bridge, consistent with the two gaseous components. Besides the Bridge's bifurcation,
it is also, on average, closer to us than the SMC \citep{bib:mfv86}. Recently, red clump (RC) and RR Lyrae stars (RRLS)
were used to estimate the SMC's line of sight
depth. \cite{bib:subramanian12} used these stars to study the three-dimensional structure of SMC and found that SMC has an average depth of
$\sim 14$~kpc. They concluded also that the NE region is closer to us and an elongation along the NE--SW direction is seen. \cite{bib:nidever13}
used RC stars to study the eastern and western regions of SMC and found that the first has a larger depth ($\sim 23$~kpc), while the second has a much
shallower depth
($\sim 10$~kpc). Some of their eastern fields showed a bimodality, with one component located at $\sim 67$~kpc and the other at $\sim 55$~kpc away from us, leading them to conclude that the closer
stellar component was stripped from SMC during the past interaction. The study of the stellar distribution of SMC shows that it has a spheroidal or slightly ellipsoidal
structure \citep{bib:subramanian12}. A stellar disk structure is also observed; however, there is a discrepancy between the gaseous and stellar disk parameters
\citep{bib:groenewegen00,bib:dobbie14,bib:subramanian14}.

	The overall picture shows that SMC possesses a stellar and a gaseous disk component, as well as a spheroidal stellar component. The eastern region and Magellanic Bridge are closer in distance to
us and show a bimodal structure. These regions
are also the ones that were probably most affected by the past interaction, which explains why they may be approaching LMC, which is located at $\sim 50$~kpc from
us. The line of sight depth estimations show a relative high dispersion from region to region, with the eastern side being deeper than the western side.

\subsection{The SMC's Magnetic Field}

	The magnetic field of SMC was previously studied using several techniques -- optical interstellar polarization, Faraday rotation, and synchrotron
emission. The interstellar polarization is caused by the alignment of the dust grains' angular momentum with a local magnetic field. The
unpolarized light emitted by the stars become
polarized due to the dichroic extinction by dust grains. Therefore, optical polarization vectors trace the magnetic field projected in the
plane of the sky, and a polarization map can be interpreted as a
magnetic field map in that direction. The phenomenon known as Faraday rotation happens, when radiation travels through an ionized medium with a local
magnetic field. The light plane of polarization rotates, if the local magnetic field has a parallel component to the direction of propagation of the light. The incident light is decomposed in to two
parts circularly polarized with opposite rotations. The synchrotron emission is produced when charged particles are
accelerated by a magnetic field. The radiation emitted by these particles is polarized, because the acceleration is not isotropic.

	\cite{bib:mao08} used optical polarization data from \citet{bib:mf70a} and obtained an ordered magnetic field
projected in the plane of the sky of
$B_{sky}~=~(1.6~\pm~0.4)~\mu$G and a turbulent component of $2~\mu$G. \cite{bib:magalhaes09} analyzed optical polarization data from
\cite{bib:magalhaes90} and estimated an ordered sky-projected
magnetic field of $1.7~\mu$G and a turbulent field of $3.5~\mu$G. \cite{bib:mao08} also estimated the line of sight ordered magnetic field, through Faraday rotation, using data from the Australia Telescope Compact Array (ATCA).
They obtained $B_{LOS}=(0.19\pm0.06)~\mu$G, with the field pointing away from us. In addition, there are several studies based on the radio continuum synchrotron emission \citep{bib:haynes86,bib:loiseau87,bib:haynes90}. Using synchrotron emission and assuming energy equipartition between the cosmic
rays and magnetic fields, \cite{bib:loiseau87} estimated the SMC's total magnetic field to be about $5~\mu$G.

	Photoelectric polarization data from \cite{bib:magalhaes90,bib:magalhaes09}, \cite{bib:schmidt70,bib:schmidt76}, and \cite{bib:mf70a,bib:mf70b}
suggested that SMC has an ordered magnetic field aligned with the Magellanic Bridge. Polarized synchrotron emission, despite its weak nature in the SMC, also confirmed such an alignment
\citep{bib:haynes90}. Besides the alignment with the Bridge, \cite{bib:magalhaes09} observed an alignment of the ordered magnetic field with the
SMC's Bar. The magnetic field may have played an important role in shaping the Bridge, since the interstellar magnetic field of galaxies can be
strong enough to
influence the dynamics of their gas \citep{bib:zweibel97}. Therefore, this work aims to study the structure of the magnetic field using CCD data, which is
more precise than photoelectric. We concentrated our study on the NE and Wing regions of the SMC, since
these regions may have been more affected by the last interaction with LMC. We also have some data of the Magellanic Bridge, from the part closer to the SMC, a structure
probably formed due the interaction of the SMC--LMC system.

	The paper is structured as follows. In Section \ref{sec:data}, we describe the observational data, the reduction process, and how we built the
catalog. The estimates for the foreground polarization are presented in Section \ref{sec:foreground}. We analyze the polarization patterns for each
field observed in Section \ref{sec:Ppatterns}. The magnetic field geometry is discussed in Section \ref{sec:Bgeometry}.
The alignment of the ordered magnetic field with respect to the Magellanic Bridge is discussed in Section \ref{sec:alignment}. We estimate the magnetic
field strength in Section \ref{sec:Bintensity}.
Finally, in Section \ref{sec:conclusion}, we summarize our results and the main conclusions.


\section{Observations and Data Reduction}\label{sec:data}

\subsection{Observational Data}

	The observational data were taken at Cerro Tololo Inter-American Observatory (CTIO) over the course of five nights in 1992 November 13--17, applying the optical polarization technique and using the
$1.5$~m telescope. The instrumental setup consisted of a half-wave retarder plate followed by a calcite,
$V$ filter, and the detector. This setup gives us, for each star, two images with
orthogonal polarizations. At least two sets of images, taken with the half-wave retarder plate positioned at different
angles, are necessary for each object/field, to obtain the linear Stokes' parameters $Q$
and $U$ \citep{bib:serkowski74}. Some standard stars polarization were measured using 16 images (see Table \ref{tab:standards}).
Further details about the instrumental technique can be found in \cite{bib:magalhaes96}. A Tek1K-1 1k~x~1k CCD, with a plate scale of
$0''.434$/pixel, read noise of $8.0602$~e$^{-}$~rms, and gain of $9.005$~e$^{-}$/ADU was used.

	During the observational run, bias and flat-field images were taken to correct instrumental noise. One non-polarized standard star was observed, in
order to check whether there was any instrumental polarization or not. Also, three polarized standard stars were observed, in order to get the
conversion for the polarization angles ($\theta$)\footnote[6]{The polarization angle $\theta$ is most common refereed as the position angle (PA).
Throughout this paper we use the notation $\theta$ instead of PA.} into the equatorial system. Twenty-eight fields of 8~x~8~arcmin size, situated at the NE and Wing sections of the SMC as well as
the Magellanic Bridge, were observed. We decided to concentrate our data efforts on these regions for the reasons mentioned previously, relating to the last interaction with the LMC, about $0.2$~Gyr
ago \citep{bib:westerlund91}.

\begin{figure*}[!htb]\centering
\includegraphics[width=.9\textwidth]{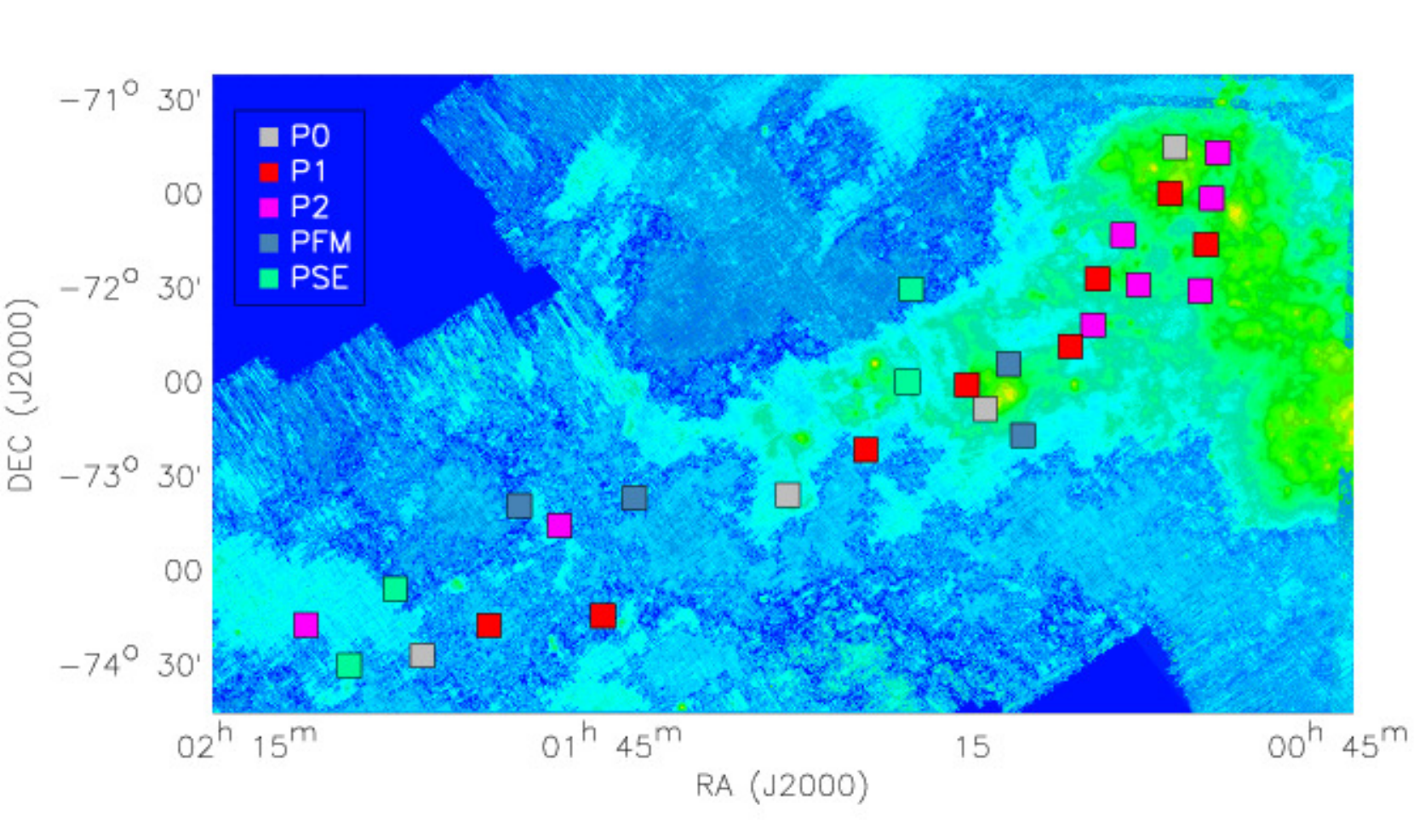}
\caption{Observed SMC fields overlapped with a {\it Spitzer}/MIPS image at $160~\mu$m \citep{bib:gordon11}. The square size represents the real
FOV. The fields with no polarization trend are labeled as P0; the fields with one polarization trend are
labeled as P1; the fields with two polarization trends are labeled as P2; the fields that were filtered by
magnitudes are labeled as PFM; and the fields that had stars excluded are labeled as PSE. The classification for the fields is introduced at Section
\ref{sec:Ppatterns}.}
\label{fig:fields}
\end{figure*}

	The date of the observation, number of positions the retarder plate was rotated, and integration time for the standard stars are presented in Table
\ref{tab:standards}. The coordinates and date of the observation for the 28 fields are presented in Table \ref{tab:fields}. For the fields, the
half-wave retarder plate was rotated four
times and the integration time was $300$~s per frame. Figure \ref{fig:fields} shows the positions of these fields.

\begin{table}[!htb]\centering
\caption{Observations of standard stars}
\begin{threeparttable}
\begin{tabular}{ccccc}\toprule\toprule

HD & Date & No.\tnote{a} & IT\tnote{b} & Standard \\
 & & & (s) & \\\midrule

9540    & 1992 Nov 13 & 16 & 20 & Non-polarized \\
283812  & 1992 Nov 13 & 16 & 8 & Polarized \\
23512   & 1992 Nov 14 & 16 & 3 & Polarized \\
23512   & 1992 Nov 15 & 4 & 3 & Polarized \\
23512   & 1992 Nov 16 & 4 & 3 & Polarized \\
23512   & 1992 Nov 17 & 4 & 4 & Polarized \\
298383  & 1992 Nov 17 & 4 & 6 & Polarized \\\bottomrule

\end{tabular}

\begin{tablenotes}
{\footnotesize
\item[a] Number of images.
\item[b] Integration time in each image.
}
\end{tablenotes}

\label{tab:standards}
\end{threeparttable}
\end{table}

\begin{table}[!htb]\centering
\caption{Observations of SMC fields}
\begin{threeparttable}
\begin{tabular}{cccc}\toprule\toprule

SMC\tnote{a} & R.A.\tnote{b} & Decl.\tnote{b} & Date \\
 & (h:m:s) & ($\arcdeg$:$\arcmin$:$\arcsec$) & \\\midrule

01 & 01:00:50.2 &-71:51:55.2 & 1992 Nov 13 \\
02 & 01:00:49.8 &-72:06:56.2 & 1992 Nov 13 \\
03 & 01:00:49.4 &-72:21:55.2 & 1992 Nov 13 \\
04 & 01:00:49.0 &-72:36:56.2 & 1992 Nov 13 \\
05 & 01:03:48.9 &-71:51:57.0 & 1992 Nov 13 \\
06 & 01:03:48.5 &-72:06:58.0 & 1992 Nov 13 \\
07 & 01:05:16.9 &-72:37:27.8 & 1992 Nov 14 \\
08 & 01:06:46.7 &-72:21:59.8 & 1992 Nov 14 \\
09 & 01:08:15.6 &-72:36:59.7 & 1992 Nov 14 \\
10 & 01:08:15.1 &-72:52:00.7 & 1992 Nov 14 \\
11 & 01:09:44.2 &-72:59:31.7 & 1992 Nov 14 \\
12 & 01:12:41.7 &-73:29:33.7 & 1992 Nov 14 \\
13 & 01:14:11.9 &-73:07:03.7 & 1992 Nov 14 \\
14 & 01:15:40.6 &-73:22:04.7 & 1992 Nov 14 \\
15 & 01:17:10.2 &-73:14:35.8 & 1992 Nov 14 \\
16 & 01:21:39.4 &-72:44:39.1 & 1992 Nov 17 \\
17 & 01:21:38.2 &-73:14:40.1 & 1992 Nov 15 \\
18 & 01:24:35.8 &-73:37:11.4 & 1992 Nov 16 \\
19 & 01:30:32.4 &-73:52:17.3 & 1992 Nov 15 \\
20 & 01:42:26.8 &-73:52:26.9 & 1992 Nov 15 \\
21 & 01:45:23.3 &-74:30:00.8 & 1992 Nov 15 \\
22 & 01:48:23.7 &-74:00:03.7 & 1992 Nov 15 \\
23 & 01:51:22.8 &-73:52:36.7 & 1992 Nov 16 \\
24 & 02:01:46.6 &-74:15:17.8 & 1992 Nov 15 \\
25 & 02:00:15.8 &-74:37:45.2 & 1992 Nov 17 \\
26 & 02:06:13.1 &-74:37:52.9 & 1992 Nov 15 \\
27 & 02:09:12.8 &-74:22:54.3 & 1992 Nov 15 \\
28 & 01:54:37.9 &-74:30:20.1 & 1992 Nov 15 \\\bottomrule

\end{tabular}

\begin{tablenotes}
{\footnotesize
\item[a] Field's label.
\item[b] Coordinates in J2000.
}
\end{tablenotes}

\label{tab:fields}
\end{threeparttable}
\end{table}

\subsection{Reduction Process}

	The data were reduced using the software IRAF\footnote[7]{IRAF is distributed by the National Optical Astronomy Observatories, which are operated
by the Association of Universities for Research in Astronomy, Inc., under cooperative agreement with the National Science Foundation.}, more specifically the NOAO and PCCD packages. This last package was developed by the Polarimetric Group from
IAG/USP \citep{bib:pereyra00} to deal with polarimetric data.

	We eliminated instrumental noise applying overscan, bias, and flat-field corrections. Once the images were processed, the Stokes $Q$
and $U$ parameters were obtained performing aperture photometry of ordinary and extraordinary images for each stellar object in the field. This was done in an automated manner,
firstly looking for all objects with a stellar profile, then performing the photometry for each of these, finally using the set of image pairs the Stokes
parameters $Q$ and $U$ were obtained. More details about the polarimetric tasks can be found in \cite{bib:pereyra00}.

	A polarization of $P=(0.035\pm0.027)\%$ was obtained for the non-polarized star HD9540, a value within $1\sigma$
of the published one \citep{bib:heiles00}. The instrumental polarization was therefore found to be negligible. For the polarized stars, we found
polarization intensities, with values within $3\sigma$ of ones published by \cite{bib:hsu82}, \cite{bib:bastien88}, and \cite{bib:wisconsin}. One exception to this was HD298383, that lay within
$\sim 3.4\sigma$. The discrepancy for this measurement is due to the very small error of the published value by
\cite{bib:tapia88}. The good agreement indicates that
the instrument measured the polarizations precisely. Lastly, using the polarization angles from the polarized standard stars, we estimated the $\Delta \theta$
conversion for the equatorial system. The standard deviation for $\Delta \theta$ is $1^{\circ}.8$,
indicating a good determination of the conversion factor.

	After the reduction process, we carefully analyzed the stars that had suspiciously high polarizations. Stars too
bright, too faint, or with badly centered coordinates tended to have such high polarizations. Since the measurements for these stars are not reliable,
we discarded them.

\subsection{Polarimetric Catalog}

	The polarimetric catalog contains 7207 stars and was constructed using stars with $P/\sigma_P > 3$. This new catalog is a huge
improvement in the number of stars compared to previous polarimetric catalogs for the SMC \citep{bib:mf70a,bib:schmidt76}. It consists of positions,
polarization intensity and its associated error, polarization angle, and $V$ magnitude and the corresponding error. A short version of the
polarimetric catalog, for the observed and intrinsic polarizations, can be seen in the Appendix \ref{sec:appendixa}.

	In order to obtain the position of the stars in right ascension (R.A.) and declination (decl.), we made use of Digital Sky Survey (DSS) images. These
images had the same center and size as our fields. Then, using a reference star, we performed a transformation
from pixel to astronomical coordinates. The average errors for the positions are $\sigma_{R.A.}=0.25~s$ and $\sigma_{decl.}=1.0''$. About $12\%$ of the stars have repeated coordinates, due to the close
proximity of a bright and faint star in the images. This occurs in the astrometry procedure when the faint stars are re-centered at the bright star positions, due to software limitations. We nevertheless
decided to keep those stars in the catalog since the exact positions are not important for the further magnetic field study.

	The $V$ magnitude ($m_V$) was obtained through summing up the flux of the image pair using one frame. In order to convert these
magnitudes into the $UBV$ system, we made use of published catalogs. Table \ref{tab:Vmag} presents the
references for these catalogs and the number of stars used to perform the calibration for each field. For some fields, we used more than one star to estimate
the $\Delta m_V$ calibration parameter. For others, using the same star we looked for its $m_V$ in different catalogs. The average error for $m_V$ is $\sigma_{m_V}=0.13$~mag, which was obtained by averaging the individual errors. These errors take into account the
photon noise error and the dispersion in the determination of $\Delta m_V$.

\begin{table*}[!htb]\centering
\caption{Photometric calibration}
\begin{threeparttable}
\begin{tabular}{ccl}\toprule\toprule

SMC\tnote{a} & No. of Stars\tnote{b} & Reference\tnote{c}  \\\midrule

01-19 & 4 & \cite{bib:massey02} \\
20-21 & 1 & \cite{bib:ardeberg77} , \cite{bib:isserstedt78}, and \cite{bib:gsc07} \\
22 & 1 & \cite{bib:isserstedt78} and \cite{bib:ardeberg80} \\
23 & 2 & \cite{bib:gsc07} and \cite{bib:sanduleak89} \\
24 & 1 & \cite{bib:gsc07} and \cite{bib:perie91} \\
25 & 1 & \cite{bib:gsc07} \\
26-27 & 3 & \cite{bib:demers91} \\
28 & 2 & \cite{bib:gsc07} \\\bottomrule

\end{tabular}

\begin{tablenotes}
{\footnotesize
\item[a] SMC field labels.
\item[b] Number of stars used.
\item[c] Reference for the catalogs used to make the $m_V$ calibration.
}
\end{tablenotes}

\label{tab:Vmag}
\end{threeparttable}
\end{table*}

\begin{figure}[!htp]\centering
\includegraphics[width=1.\columnwidth]{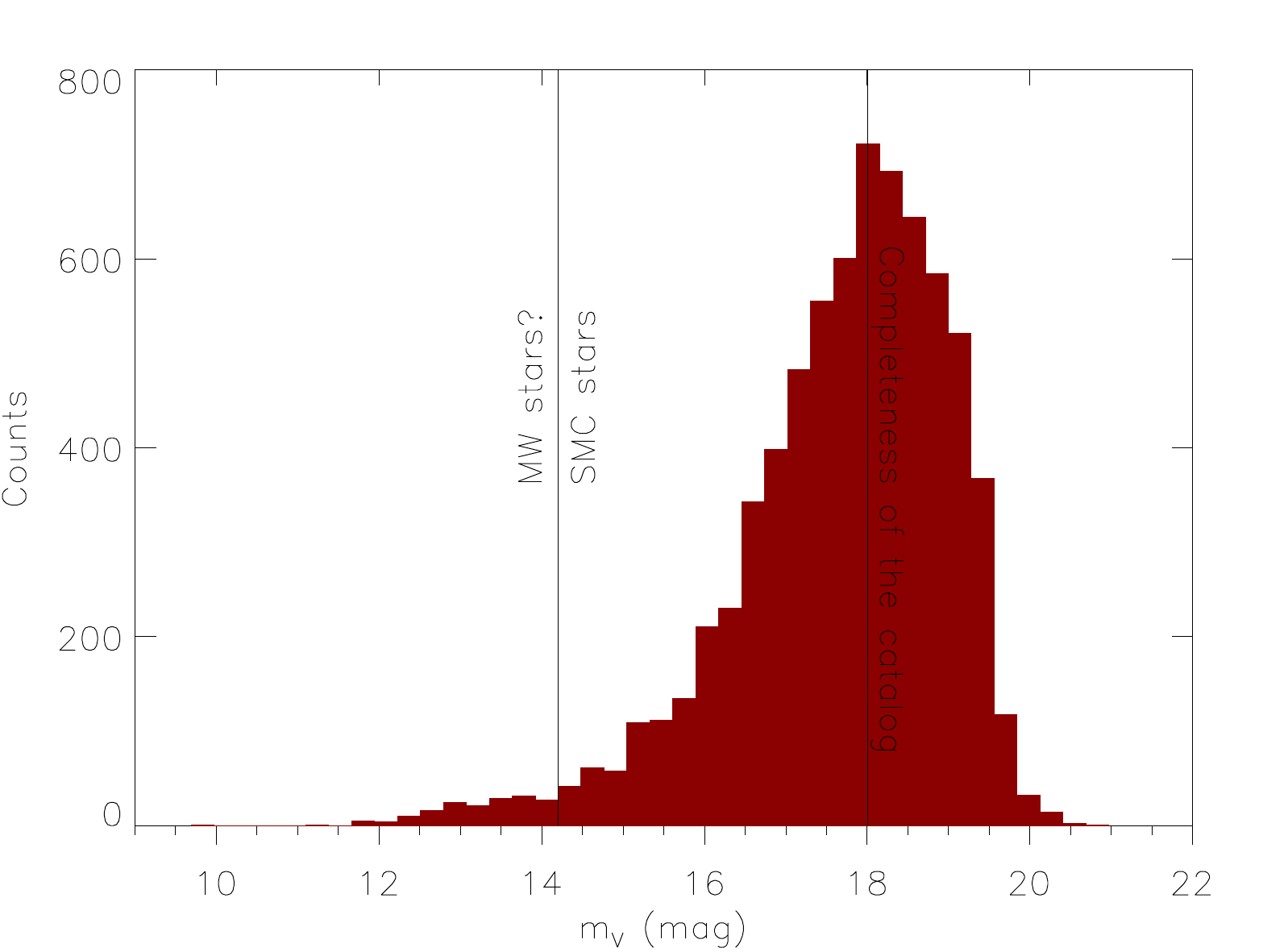}
\caption{$V$ magnitude histogram of the catalog stars. The vertical line at $m_V~=~14.2$~mag separates the SMC stars from the foreground candidates. It is also
indicated the limit magnitude ($m_{V,\text{lim}}~=~18.00 \pm 0.28$~mag) of the catalog.}
\label{fig:hmV}
\end{figure}

	To evaluate the completeness of the catalog, we constructed a
histogram with bin sizes of $1$~mag. The completeness or limit magnitude was determined by the histogram's peak. Afterwards, we evaluated the standard deviation in the bin of the limiting magnitude. A new
histogram with bin sizes equal to the standard deviation ($\sigma_{m_V} = 0.28$~mag) was constructed, which can be seen in Figure \ref{fig:hmV}.
Finally, the new limiting magnitude was defined as the new peak of the histogram. Using this procedure, the catalog's
completeness is $m_V = 18.00$~mag. We suspect that most of the stars with $m_V \lesssim 14$~mag
belong to the Galaxy. In Section \ref{sec:foreground} we further discuss this assumption.

\section{Foreground Polarization}\label{sec:foreground}

	The light from the SMC's member stars has to travel through the Galaxy's ISM. Therefore, part of the polarization we measure
is due to the MW's dust. The SMC's interstellar polarization is obtained by subtracting the foreground polarization from the Galaxy, which is
estimated in this section. 

	\cite{bib:schmidt76} divided the SMC into five regions (see Figure 2 in this paper) and estimated the foreground polarization for these regions using
foreground stars.
\cite{bib:rodrigues97} re-estimated the foreground polarization for each of
these regions and obtained more precise values. In our analysis, we used the same regions from \cite{bib:schmidt76} and added a sixth region between the
SMC's body and the Magellanic Bridge, which we call region IV--V. One of the motivations to estimate the foreground polarization from
our data is to be
able to subtract the Galactic foreground in that new region.

        Our polarimetric data is spread along the NE and Wing regions of SMC and the Magellanic Bridge. Therefore it does not include objects
from Schmidt (1976)'s region I and that is why we do not have an estimate for the Galactic foreground there. Region II corresponds to the NE part of the SMC. Regions III and IV represent the SMC's
Wing, with the former located nearer to the SMC's Bar. Region IV--V is located between regions IV and V, the latter being located furthest relative to the SMC and at the Magellanic Bridge.

	Our catalog has both SMC and Galaxy stars. Thus, one way to estimate the foreground polarization is to determine which stars
are located in the Galaxy and use their polarizations as a measure for the foreground. Below we estimate a magnitude threshold for SMC stars.

	The distance modulus relation is
\begin{equation}
m_V - M_V = 5~log~d - 5 + A_V,
\label{eq:dmodulus}
\end{equation}
where $m_V$ is the apparent magnitude in the $V$ band; $M_V$ is the absolute magnitude in the $V$ band; $d$ is the distance in pc; and $A_V$ is the
MW foreground extinction in the $V$ band.

	The SMC's distance is $d=(63\pm1)~$kpc \citep{bib:cioni00}, which was obtained using the tip of the red giant branch method. Recently,
\cite{bib:scowcroft15} obtained $d=(62\pm0.3)$~kpc, using {\it Spitzer} observations of classical Cepheids, the new measurement agrees with the distance we assumed. In order to estimate the average foreground extinction, one can use the average
foreground color excess, $E(B~-~V)~\simeq~0.05~$mag \citep{bib:bessell91}, together with the average extinction law for the MW, $A_V \simeq 3.2E(B-V)$
\citep{bib:seaton79}, thus we get $A_V \simeq 0.2~$mag. Hence, the distance modulus to the SMC is $m_V-M_V \simeq 19.2$~mag.

	The brightest giant and super-giant stars have $M_V~<~-~5$~mag, which translates into $m_V \lesssim 14.2~$mag when considering the previously mentioned distance modulus for the SMC. \cite{bib:massey02}'s catalog focused on a precise magnitude calibration for bright
stars from the SMC, to complement other catalogs focused on faint stars. This catalog has 84995 stars with $m_V \lesssim 18~$mag, from which just
$3.5\%$ have $m_V \lesssim 14.2$~mag. Therefore, considering that all stars with $m_V < 14.2$~mag are Galactic objects, there is a possible inclusion of only a small number of SMC members.

	In addition, all the stars used by \cite{bib:schmidt76} and \cite{bib:rodrigues97} to estimate the foreground polarization have $m_V \lesssim 14.2~$mag.
Considering this fact and the argument above, it is a reasonable approximation to characterize the stars from our catalog with magnitudes below this value
as foreground star candidates. Using this limit we would be excluding from the foreground sample just the Galactic late-type dwarfs.

        A second filter was applied to select objects representative of foreground polarization. We excluded all objects with polarizations higher than $1\%$ and differing
from the average, of the stars with $m_V < 14.2$~mag, by a factor larger than $2\sigma$. This procedure may remove stars with intrinsic polarization.

	The steps to separate the foreground stars and estimate the foreground polarization were the following:

\begin{enumerate}

\item Take the stars with $m_V<14.2$~mag and $P<1\%$;

\item Obtain the Stokes parameters $Q$ and $U$ and its associated error for the selected stars;

\item Divide the stars into the \cite{bib:schmidt76}'s regions and region IV--V;

\item Obtain the uncertainty weighted average Stokes parameters for each region and its standard deviation;

\item Exclude the stars with Stokes parameters deviating more than $2\sigma$ from the average;

\item Using the remaining stars, the Stokes parameters are re-computed using an uncertainty-weighted average. We estimated the error of the Galactic
polarization in each region as the mean standard deviation,
because the error from the uncertainty weighted average is usually underestimated;

\item Finally, from the Stokes parameters, we obtain the polarization, its angle, and the error for each region.

\end{enumerate}

	Table \ref{tab:foreground} shows our estimate for the foreground along side that of \cite{bib:schmidt76} and \cite{bib:rodrigues97}. It can be seen that our estimates are in good agreement with the
two previously mentioned works. We obtained smaller errors due to the higher precision of our CCD measurements compared with previous
photoelectric polarimetry. In addition, the ratio of foreground polarization to foreground extinction is smaller than $3\%$ mag$^{-1}$ in all regions, as expected for the Galaxy
\citep{bib:whittet92}. These facts provide a reasonable degree of confidence in our estimates. We used
our values to subtract the foreground polarization from the stars of our catalog. The foreground subtraction was done via the Stokes parameters.

\begin{table*}[!htp]\centering
\caption{Foreground polarization for each region}
\begin{threeparttable}
\begin{tabular}{cccccccc}\toprule\toprule

\multirow{2}{*}{Region\tnote{a}} & $P_{\text{for,s}}$\tnote{b} & $\theta_{\text{for,s}}$\tnote{c} & $P_{\text{for,r}}$\tnote{b} &
$\theta_{\text{for,r}}$\tnote{c} & $P_{\text{for,lg}}$\tnote{b} & $\theta_{\text{for,lg}}$\tnote{c} & $N_{for,lg}$\tnote{d}\\
 & (\%) & (deg) & (\%) & (deg) & (\%) & (deg) & \\\midrule
 & \multicolumn{2}{c}{\cite{bib:schmidt76}} & \multicolumn{2}{c}{\cite{bib:rodrigues97}} & \multicolumn{3}{c}{This work}\\\midrule

I & $0.37 \pm 0.15$ & 111 & $0.47 \pm 0.09$ & 113.6 & -- & -- & -- \\
II & $0.27 \pm 0.15$ & 123 & $0.30 \pm 0.08$ & 124.1 & $0.316\pm0.016$ & 111.6 & 75 \\
III & $0.06 \pm 0.09$ & 139 & $0.17 \pm 0.05$ & 145.0 & $0.124\pm0.024$ & 136.8 & 12 \\
IV & $0.14 \pm 0.12$ & 125 & $0.22 \pm 0.05$ & 124.2 & $0.162\pm0.020$ & 125.6 & 16 \\
IV--V & -- & -- & -- & -- & $0.138\pm0.048$ & 102.7 & 15 \\
V & $0.16 \pm 0.12$ & 93 & $0.27 \pm 0.05$ & 95.6 & $0.301\pm0.028$ & 91.1 & 20 \\\bottomrule

\end{tabular}

\begin{tablenotes}
{\footnotesize
\item[a] Regions's label.
\item[b] Foreground polarization intensity ($P_{\text{for,X}}$).
\item[c] Foreground polarization angle ($\theta_{\text{for,X}}$).
\item[d] Number of stars used for the foreground estimation by this work.
}
\end{tablenotes}

\label{tab:foreground}
\end{threeparttable}
\end{table*}

\begin{figure*}[!htp]\centering
\includegraphics[width=.9\textwidth]{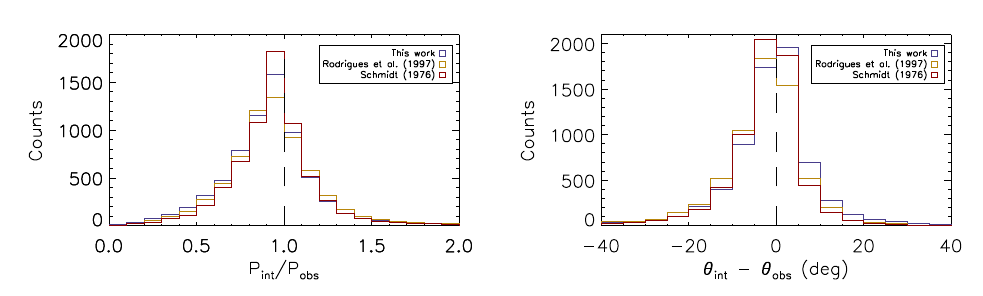}
\caption{Histograms of $P_{int}/P_{obs}$ and ($\theta_{int}-\theta_{obs}$). The red histogram is obtained after \cite{bib:schmidt76} foreground subtraction, the
orange histogram is obtained after \cite{bib:rodrigues97} foreground subtraction, and the blue histogram is obtained after foreground subtraction from this
work. The dashed lines represent no foreground.}
\label{fig:foreground}
\end{figure*}

	Figure \ref{fig:foreground} shows histograms for the ratio between the intrinsic and observed polarizations ($P_{int}/P_{obs}$) and the
difference between them ($\theta_{int} - \theta_{obs}$). The foreground subtraction can
modify the polarization intensities up to $50\%$ from the observed value. Likewise, the polarization angles can shift up to $20^{\circ}$. There is no large difference between the different foreground
corrections.

\section{Polarization Patterns in the SMC Fields}\label{sec:Ppatterns}

	With the aim of analyzing the magnetic field geometry, we studied the polarization preferential direction (PD) for each one of the 28 fields. We used stars with intrinsic polarization $P/\sigma_P >
3$, subtracting the foreground using the estimate obtained in this work. The number of objects in each field is shown in Table \ref{tab:ism}.

	In order to obtain the PDs from $\theta$, we performed Gaussian fits to the $\theta$ histograms. To find the
polarization intensity ($P$) for the PDs, we took the field stars with $\theta$ within $3\sigma$ of the mean $\theta$ obtained by the
Gaussian fits. The polarization intensity for the PDs was estimated from the median of the polarization for these stars. We used the median instead
of the average because it is a better indication of the central trend for asymmetrical distributions. This procedure was not
done for fields with a random distribution in $\theta$.

	Our fields showed different distributions for the $\theta$ histograms, therefore we classified the fields into five different groups: P0, P1, P2,
PFM, and PSE. Table \ref{tab:ism} presents the classification for each field. The next Sections explain what they are and how we dealt with them. In
order to test in a more robust way our field's classification, we performed the $F$-test on our sample. The results were consistent with our previous
classisfication. More details about the $F$-test procedure can be found in the
Appendix \ref{sec:appendixb}. Plots
of the polarization map, $\theta$ vs $m_V$, $\theta$ histogram, and polarization intensity histogram are shown in Figures \ref{fig:smc14}--4.28 (online material).

\figsetstart

\figsetgrpstart
\figsetgrpnum{4.1}
\figsetplot{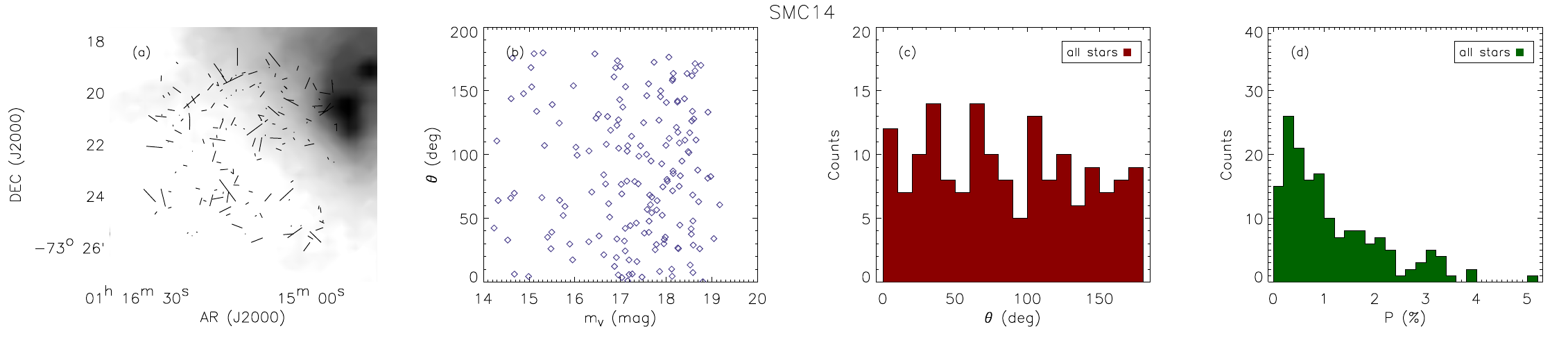}
\label{fig:smc14}
\figsetgrpnote{Plots for SMC14, a field with no PD (labeled as P0). Panel (a) shows a polarization map overlapped with a {\it Spitzer}/MIPS image at $160~\mu$m \citep{bib:gordon11}; panel (b) shows a plot of $\theta$
versus magnitude; panel (c) shows a $\theta$ histogram; and panel (d)
shows a polarization intensity histogram.}
\figsetgrpend

\figsetgrpstart
\figsetgrpnum{4.2}
\figsetplot{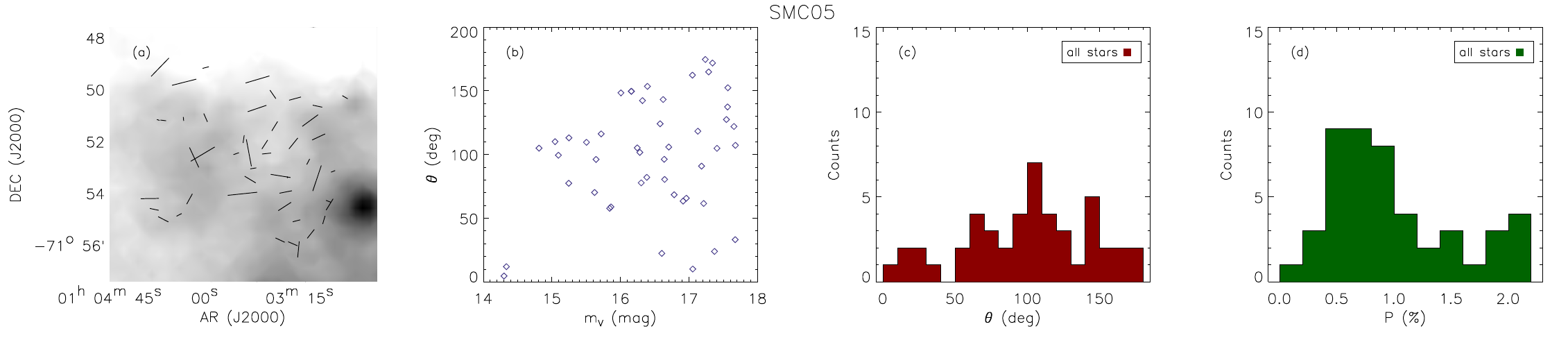}
\label{fig:smc05}
\figsetgrpnote{Plots for SMC05, a field with no PD (labeled as P0). Panel (a) shows a polarization map overlapped with a {\it Spitzer}/MIPS image at $160~\mu$m \citep{bib:gordon11}; panel (b) shows a plot of $\theta$
versus magnitude; panel (c) shows a $\theta$ histogram; and panel (d)
shows a polarization intensity histogram.}
\figsetgrpend

\figsetgrpstart
\figsetgrpnum{4.3}
\figsetplot{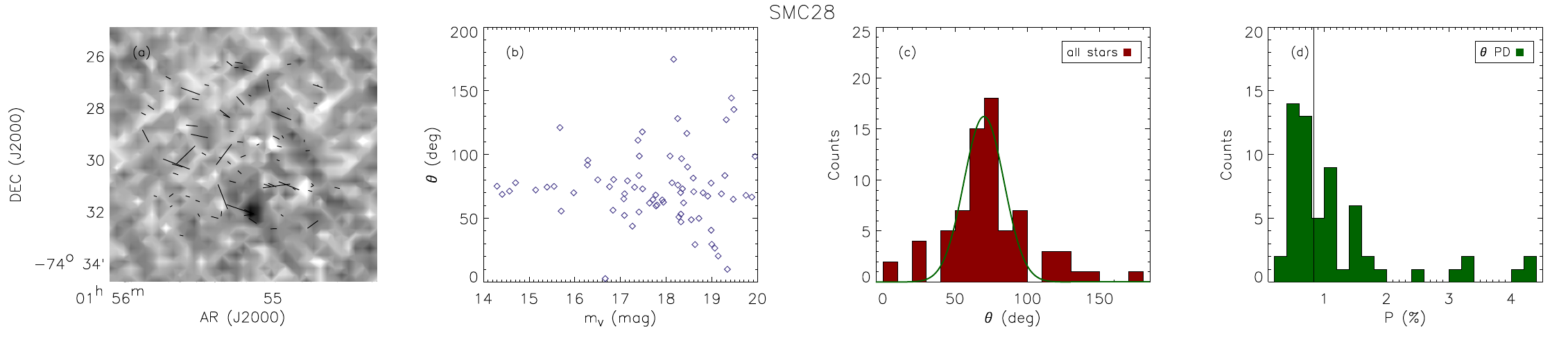}
\label{fig:smc28}
\figsetgrpnote{Plots for SMC28, a field with one PD (labeled as P1). Panel (a) shows a polarization map overlapped with a {\it Spitzer}/MIPS image at $160~\mu$m \citep{bib:gordon11}; panel (b) shows a plot of $\theta$
versus magnitude; panel (c) shows a $\theta$ histogram, the green line represents the Gaussian fit; and panel (d)
shows a polarization intensity histogram, the vertical line on this plot represents the median value.}
\figsetgrpend

\figsetgrpstart
\figsetgrpnum{4.4}
\figsetplot{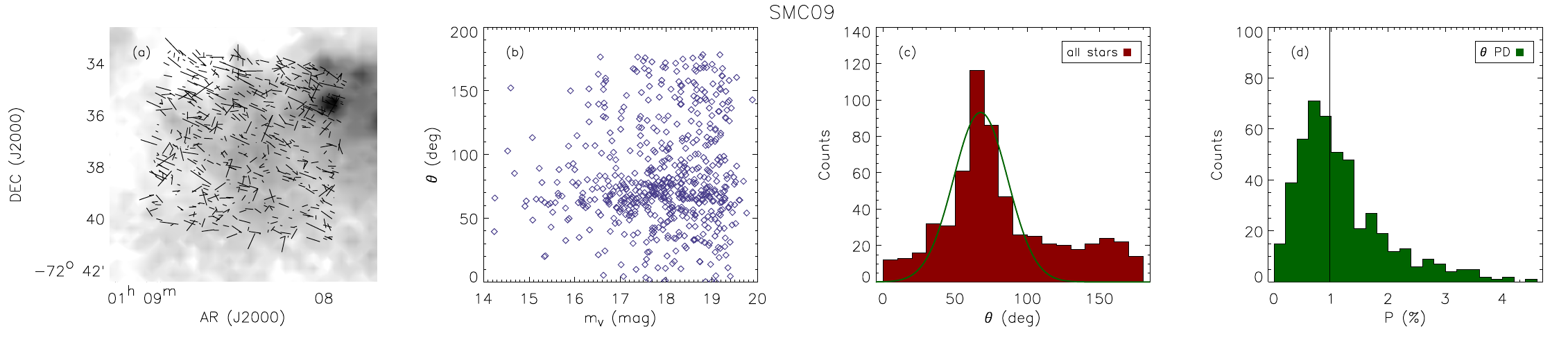}
\label{fig:smc09}
\figsetgrpnote{Plots for SMC09, a field with one PD (labeled as P1). Panel (a) shows a polarization map overlapped with a {\it Spitzer}/MIPS image at $160~\mu$m \citep{bib:gordon11}; panel (b) shows a plot of $\theta$
versus magnitude; panel (c) shows a $\theta$ histogram, the green line represents the Gaussian fit; and panel (d)
shows a polarization intensity histogram, the vertical line on this plot represents the median value.}
\figsetgrpend

\figsetgrpstart
\figsetgrpnum{4.5}
\figsetplot{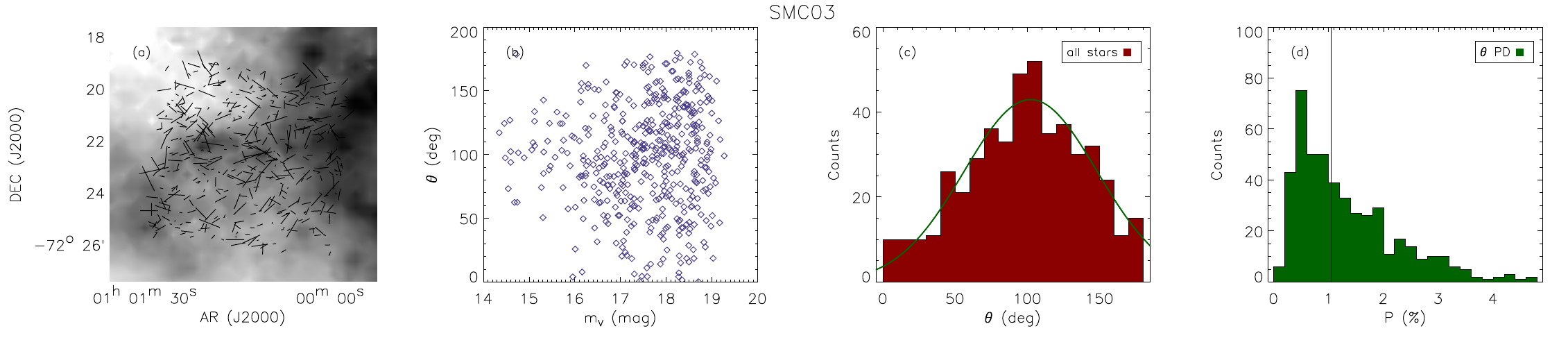}
\label{fig:smc03}
\figsetgrpnote{Plots for SMC03, a field with one PD (labeled as P1). Panel (a) shows a polarization map overlapped with a {\it Spitzer}/MIPS image at $160~\mu$m \citep{bib:gordon11}; panel (b) shows a plot of $\theta$
versus magnitude; panel (c) shows a $\theta$ histogram, the green line represents the Gaussian fit; and panel (d)
shows a polarization intensity histogram, the vertical line on this plot represents the median value.}
\figsetgrpend

\figsetgrpstart
\figsetgrpnum{4.6}
\figsetplot{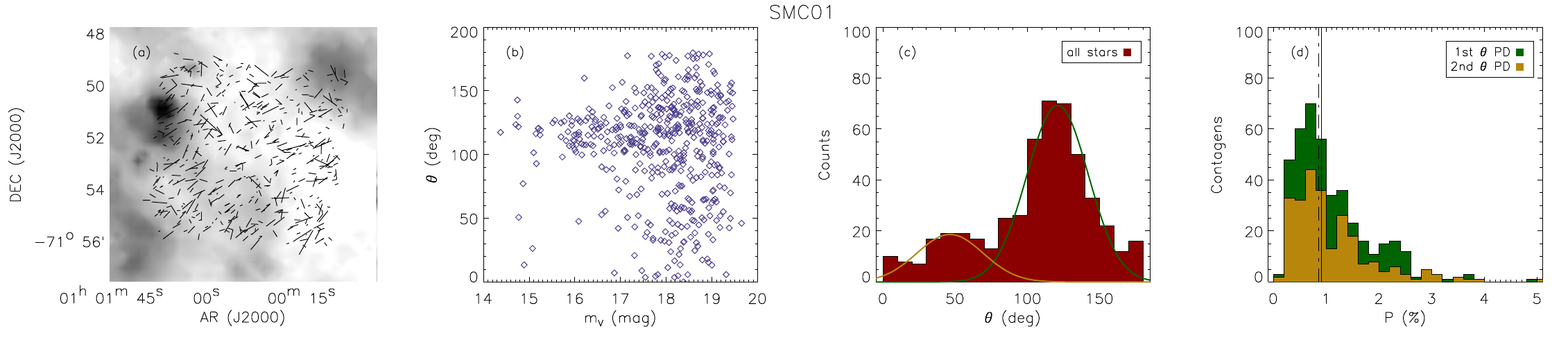}
\label{fig:smc01}
\figsetgrpnote{Plots for SMC01, a field with two PDs (labeled as P2). Panel (a) shows a polarization map overlapped with a {\it Spitzer}/MIPS image at $160~\mu$m \citep{bib:gordon11}; panel (b) shows a plot of $\theta$
versus magnitude; panel (c) shows a $\theta$ histogram, the green and gold lines represent the Gaussian fits for the first
and second PDs, respectively; and panel (d)
shows a polarization intensity histogram, the vertical lines on this plot represents the median value for the first (solid line) and second PD
(solid-dotted line).}
\figsetgrpend

\figsetgrpstart
\figsetgrpnum{4.7}
\figsetplot{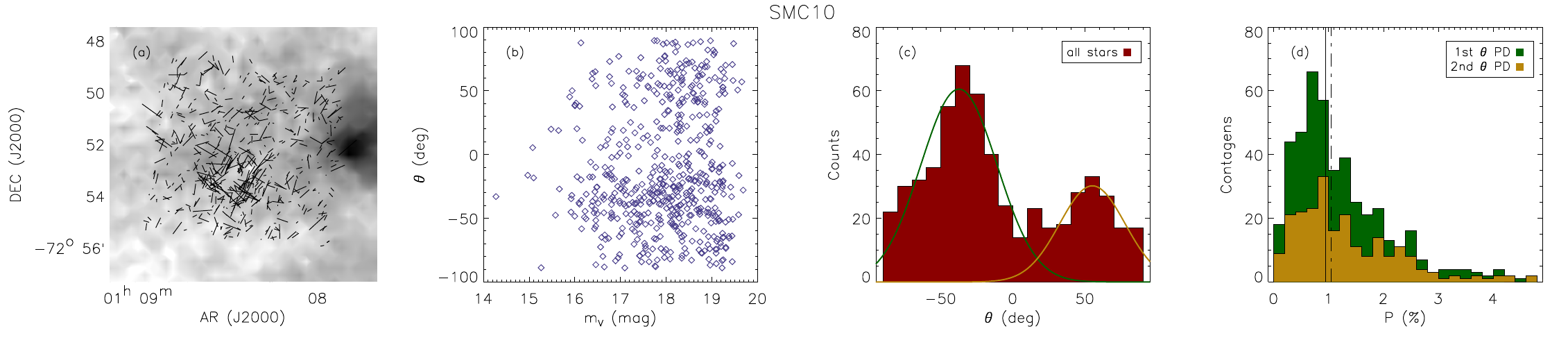}
\label{fig:smc10}
\figsetgrpnote{Plots for SMC10, a field with two PDs (labeled as P2). Panel (a) shows a polarization map overlapped with a {\it Spitzer}/MIPS image at $160~\mu$m \citep{bib:gordon11}; panel (b) shows a plot of $\theta$
versus magnitude; panel (c) shows a $\theta$ histogram, the green and gold lines represent the Gaussian fits for the first
and second PDs, respectively; and panel (d)
shows a polarization intensity histogram, the vertical lines on this plot represents the median value for the first (solid line) and second PD
(solid-dotted line).}
\figsetgrpend

\figsetgrpstart
\figsetgrpnum{4.8}
\figsetplot{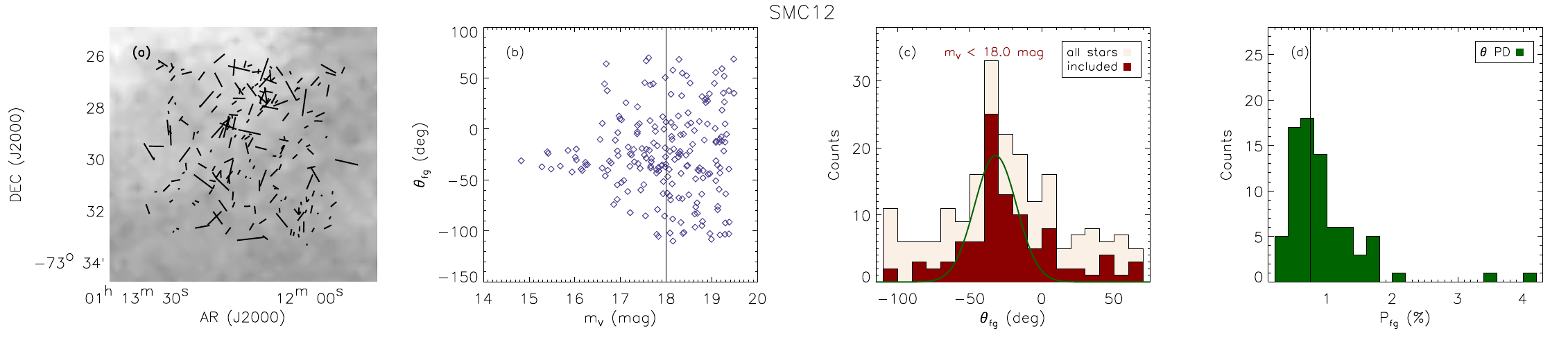}
\label{fig:smc12}
\figsetgrpnote{Plots for SMC12, a field for which we had to filter the stars by magnitudes to fit the polarization PD (labeled as PFM). Panel
(a) shows a polarization map overlapped with a {\it Spitzer}/MIPS image at $160~\mu$m \citep{bib:gordon11}; panel (b) shows a plot of $\theta$ versus magnitude, the vertical line demarcates the magnitude cut; panel (c)
shows a $\theta$ histogram, the green line represents the Gaussian fit; and panel (d)
shows a polarization intensity histogram, the vertical line on this plot represents the median value.}
\figsetgrpend

\figsetgrpstart
\figsetgrpnum{4.9}
\figsetplot{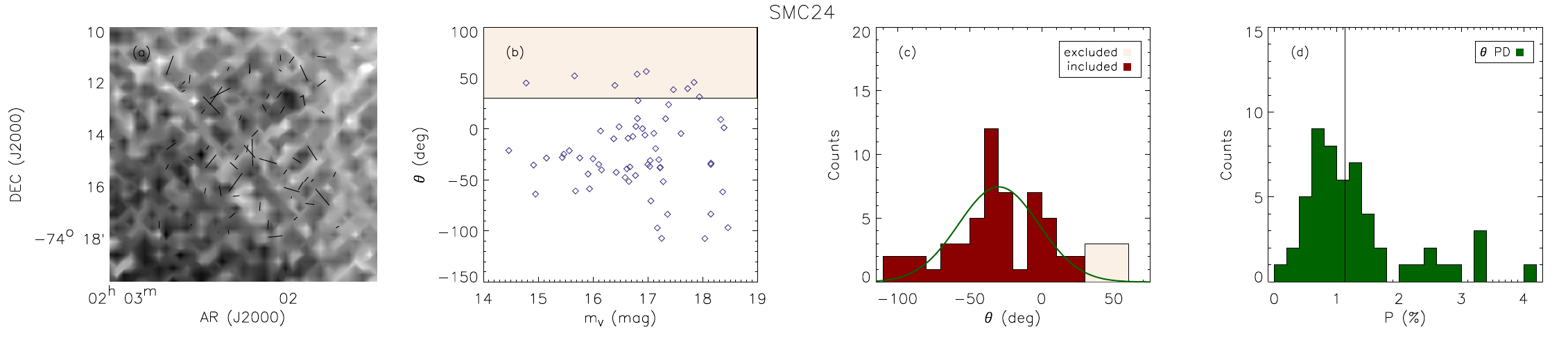}
\label{fig:smc24}
\figsetgrpnote{Plots for SMC24, a field for which we had to exclude some stars to fit the polarization PD (labeled as PSE). Panel (a) shows a
polarization map overlapped with a {\it Spitzer}/MIPS image at $160~\mu$m \citep{bib:gordon11}; panel (b) shows a plot of $\theta$ versus magnitude, the stars in the light pink area were not considered for the
Gaussian fit; panel (c) shows a $\theta$ histogram, the green line represents the Gaussian fit; and panel (d)
shows a polarization intensity histogram, the vertical line on this plot represents the median value.}
\figsetgrpend

\figsetgrpstart
\figsetgrpnum{4.10}
\figsetplot{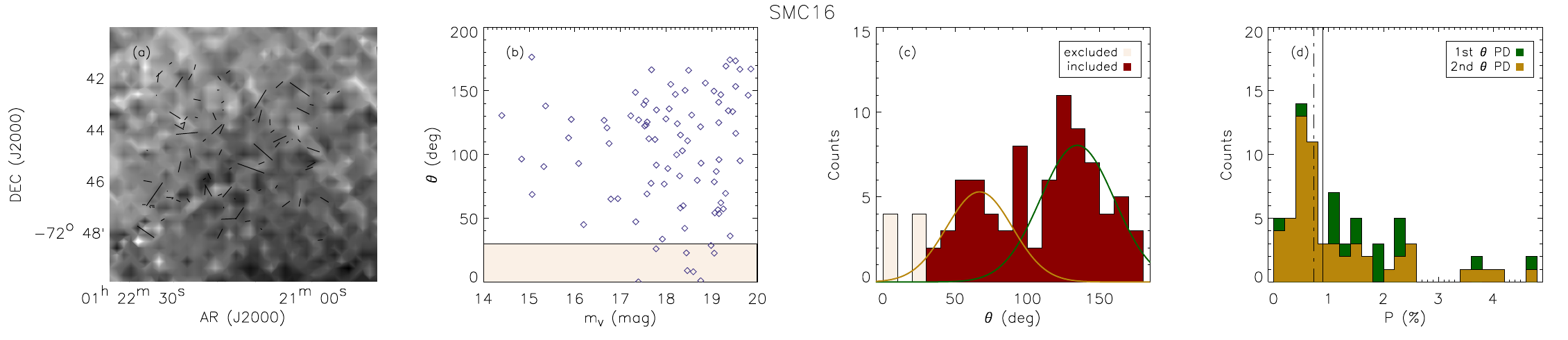}
\label{fig:smc16}
\figsetgrpnote{Plots for SMC16, a field for which we had to exclude some stars to fit the polarization PD (labeled as PSE). Panel (a) shows a
polarization map overlapped with a {\it Spitzer}/MIPS image at $160~\mu$m \citep{bib:gordon11}; panel (b) shows a plot of $\theta$ versus magnitude, the stars in the light pink area were not considered for the
Gaussian fit; panel (c) shows a $\theta$ histogram, the green and gold lines represent the Gaussian fits for the first
and second PDs, respectively; and panel (d)
shows a polarization intensity histogram, the vertical lines on this plot represents the median value for the first (solid line) and second PD
(solid-dotted line).}
\figsetgrpend

\figsetgrpstart
\figsetgrpnum{4.11}
\figsetplot{smc02_gf.pdf}
\label{fig:smc02}
\figsetgrpnote{Plots for SMC02, a field with two PDs (labeled as P2). Panel (a) shows a polarization map overlapped with a {\it Spitzer}/MIPS image at $160~\mu$m \citep{bib:gordon11}; panel (b) shows a plot of $\theta$
versus magnitude; panel (c) shows a $\theta$ histogram, the green and gold lines represent the Gaussian fits for the first
and second PDs, respectively; and panel (d)
shows a polarization intensity histogram, the vertical lines on this plot represents the median value for the first (solid line) and second PD
(solid-dotted line).}
\figsetgrpend

\figsetgrpstart
\figsetgrpnum{4.12}
\figsetplot{smc04_gf.pdf}
\label{fig:smc04}
\figsetgrpnote{Plots for SMC04, a field with two PDs (labeled as P2). Panel (a) shows a polarization map overlapped with a {\it Spitzer}/MIPS image at $160~\mu$m \citep{bib:gordon11}; panel (b) shows a plot of $\theta$
versus magnitude; panel (c) shows a $\theta$ histogram, the green and gold lines represent the Gaussian fits for the first
and second PDs, respectively; and panel (d)
shows a polarization intensity histogram, the vertical lines on this plot represents the median value for the first (solid line) and second PD
(solid-dotted line).}
\figsetgrpend

\figsetgrpstart
\figsetgrpnum{4.13}
\figsetplot{smc06_gf.pdf}
\label{fig:smc06}
\figsetgrpnote{Plots for SMC06, a field with one PD (labeled as P1). Panel (a) shows a polarization map overlapped with a {\it Spitzer}/MIPS image at $160~\mu$m \citep{bib:gordon11}; panel (b) shows a plot of $\theta$
versus magnitude; panel (c) shows a $\theta$ histogram, the green line represents the Gaussian fit; and panel (d)
shows a polarization intensity histogram, the vertical line on this plot represents the median value.}
\figsetgrpend

\figsetgrpstart
\figsetgrpnum{4.14}
\figsetplot{smc07_gf.pdf}
\label{fig:smc07}
\figsetgrpnote{Plots for SMC07, a field with two PDs (labeled as P2). Panel (a) shows a polarization map overlapped with a {\it Spitzer}/MIPS image at $160~\mu$m \citep{bib:gordon11}; panel (b) shows a plot of $\theta$
versus magnitude; panel (c) shows a $\theta$ histogram, the green and gold lines represent the Gaussian fits for the first
and second PDs, respectively; and panel (d)
shows a polarization intensity histogram, the vertical lines on this plot represents the median value for the first (solid line) and second PD
(solid-dotted line).}
\figsetgrpend

\figsetgrpstart
\figsetgrpnum{4.15}
\figsetplot{smc08_gf.pdf}
\label{fig:smc08}
\figsetgrpnote{Plots for SMC08, a field with two PDs (labeled as P2). Panel (a) shows a polarization map overlapped with a {\it Spitzer}/MIPS image at $160~\mu$m \citep{bib:gordon11}; panel (b) shows a plot of $\theta$
versus magnitude; panel (c) shows a $\theta$ histogram, the green and gold lines represent the Gaussian fits for the first
and second PDs, respectively; and panel (d)
shows a polarization intensity histogram, the vertical lines on this plot represents the median value for the first (solid line) and second PD
(solid-dotted line).}
\figsetgrpend

\figsetgrpstart
\figsetgrpnum{4.16}
\figsetplot{smc11_gf.pdf}
\label{fig:smc11}
\figsetgrpnote{Plots for SMC11, a field with one PD (labeled as P1). Panel (a) shows a polarization map overlapped with a {\it Spitzer}/MIPS image at $160~\mu$m \citep{bib:gordon11}; panel (b) shows a plot of $\theta$
versus magnitude; panel (c) shows a $\theta$ histogram, the green line represents the Gaussian fit; and panel (d)
shows a polarization intensity histogram, the vertical line on this plot represents the median value.}
\figsetgrpend

\figsetgrpstart
\figsetgrpnum{4.17}
\figsetplot{smc13_gf.pdf}
\label{fig:smc13}
\figsetgrpnote{Plots for SMC13, a field for which we had to filter the stars by magnitudes to fit the polarization PD (labeled as PFM). Panel
(a) shows a polarization map overlapped with a {\it Spitzer}/MIPS image at $160~\mu$m \citep{bib:gordon11}; panel (b) shows a plot of $\theta$ versus magnitude, the vertical line demarcates the magnitude cut; panel (c)
shows a $\theta$ histogram, the green line represents the Gaussian fit; and panel (d)
shows a polarization intensity histogram, the vertical line on this plot represents the median value.}
\figsetgrpend

\figsetgrpstart
\figsetgrpnum{4.18}
\figsetplot{smc15_gf.pdf}
\label{fig:smc15}
\figsetgrpnote{Plots for SMC15, a field with one PD (labeled as P1). Panel (a) shows a polarization map overlapped with a {\it Spitzer}/MIPS image at $160~\mu$m \citep{bib:gordon11}; panel (b) shows a plot of $\theta$
versus magnitude; panel (c) shows a $\theta$ histogram, the green line represents the Gaussian fit; and panel (d)
shows a polarization intensity histogram, the vertical line on this plot represents the median value.}
\figsetgrpend

\figsetgrpstart
\figsetgrpnum{4.19}
\figsetplot{smc17_gf.pdf}
\label{fig:smc17}
\figsetgrpnote{Plots for SMC17, a field for which we had to exclude some stars to fit the polarization PD (labeled as PSE). Panel (a) shows a
polarization map overlapped with a {\it Spitzer}/MIPS image at $160~\mu$m \citep{bib:gordon11}; panel (b) shows a plot of $\theta$ versus magnitude, the stars in the light pink areas were not considered for the
Gaussian fit; panel (c) shows a $\theta$ histogram, the green and gold lines represent the Gaussian fits for the first
and second PDs, respectively; and panel (d)
shows a polarization intensity histogram, the vertical lines on this plot represents the median value for the first (solid line) and second PD
(solid-dotted line).}
\figsetgrpend

\figsetgrpstart
\figsetgrpnum{4.20}
\figsetplot{smc18_gf.pdf}
\label{fig:smc18}
\figsetgrpnote{Plots for SMC18, a field with one PD (labeled as P1). Panel (a) shows a polarization map overlapped with a {\it Spitzer}/MIPS image at $160~\mu$m \citep{bib:gordon11}; panel (b) shows a plot of $\theta$
versus magnitude; panel (c) shows a $\theta$ histogram, the green line represents the Gaussian fit; and panel (d)
shows a polarization intensity histogram, the vertical line on this plot represents the median value.}
\figsetgrpend

\figsetgrpstart
\figsetgrpnum{4.21}
\figsetplot{smc19_gf.pdf}
\label{fig:smc19}
\figsetgrpnote{Plots for SMC19, a field with no PD (labeled as P0). Panel (a) shows a polarization map overlapped with a {\it Spitzer}/MIPS image at $160~\mu$m \citep{bib:gordon11}; panel (b) shows a plot of $\theta$
versus magnitude; panel (c) shows a $\theta$ histogram; and panel (d)
shows a polarization intensity histogram.}
\figsetgrpend

\figsetgrpstart
\figsetgrpnum{4.22}
\figsetplot{smc20_gf.pdf}
\label{fig:smc20}
\figsetgrpnote{Plots for SMC20, a field for which we had to filter the stars by magnitudes to fit the polarization PD (labeled as PFM). Panel
(a) shows a polarization map overlapped with a {\it Spitzer}/MIPS image at $160~\mu$m \citep{bib:gordon11}; panel (b) shows a plot of $\theta$ versus magnitude, the vertical line demarcates the magnitude cut; panel (c)
shows a $\theta$ histogram, the green line represents the Gaussian fit; and panel (d)
shows a polarization intensity histogram, the vertical line on this plot represents the median value.}
\figsetgrpend

\figsetgrpstart
\figsetgrpnum{4.23}
\figsetplot{smc21_gf.pdf}
\label{fig:smc21}
\figsetgrpnote{Plots for SMC21, a field with one PD (labeled as P1). Panel (a) shows a polarization map overlapped with a {\it Spitzer}/MIPS image at $160~\mu$m \citep{bib:gordon11}; panel (b) shows a plot of $\theta$
versus magnitude; panel (c) shows a $\theta$ histogram, the green line represents the Gaussian fit; and panel (d)
shows a polarization intensity histogram, the vertical line on this plot represents the median value.}
\figsetgrpend

\figsetgrpstart
\figsetgrpnum{4.24}
\figsetplot{smc22_gf.pdf}
\label{fig:smc22}
\figsetgrpnote{Plots for SMC22, a field with two PDs (labeled as P2). Panel (a) shows a polarization map overlapped with a {\it Spitzer}/MIPS image at $160~\mu$m \citep{bib:gordon11}; panel (b) shows a plot of $\theta$
versus magnitude; panel (c) shows a $\theta$ histogram, the green and gold lines represent the Gaussian fits for the first
and second PDs, respectively; and panel (d)
shows a polarization intensity histogram, the vertical lines on this plot represents the median value for the first (solid line) and second PD
(solid-dotted line).}
\figsetgrpend

\figsetgrpstart
\figsetgrpnum{4.25}
\figsetplot{smc23_gf.pdf}
\label{fig:smc23}
\figsetgrpnote{Plots for SMC23, a field for which we had to filter the stars by magnitudes to fit the polarization PD (labeled as PFM). Panel
(a) shows a polarization map overlapped with a {\it Spitzer}/MIPS image at $160~\mu$m \citep{bib:gordon11}; panel (b) shows a plot of $\theta$ versus magnitude, the vertical line demarcates the magnitude cut; panel (c)
shows a $\theta$ histogram, the green line represents the Gaussian fit; and panel (d)
shows a polarization intensity histogram, the vertical line on this plot represents the median value.}
\figsetgrpend

\figsetgrpstart
\figsetgrpnum{4.26}
\figsetplot{smc25_gf.pdf}
\label{fig:smc25}
\figsetgrpnote{Plots for SMC25, a field with no PD (labeled as P0). Panel (a) shows a polarization map overlapped with a {\it Spitzer}/MIPS image at $160~\mu$m \citep{bib:gordon11}; panel (b) shows a plot of $\theta$
versus magnitude; panel (c) shows a $\theta$ histogram; and panel (d)
shows a polarization intensity histogram.}
\figsetgrpend

\figsetgrpstart
\figsetgrpnum{4.27}
\figsetplot{smc26_gf.pdf}
\label{fig:smc26}
\figsetgrpnote{Plots for SMC26, a field for which we had to exclude some stars to fit the polarization PD (labeled as PSE). Panel (a) shows a
 polarization map overlapped with a {\it Spitzer}/MIPS image at $160~\mu$m \citep{bib:gordon11}; panel (b) shows a plot of $\theta$ versus magnitude, the stars in the light pink areas were not considered for the
Gaussian fit; panel (c) shows a $\theta$ histogram, the green line represents the Gaussian fit; and panel (d)
shows a polarization intensity histogram, the vertical line on this plot represents the median value.}
\figsetgrpend

\figsetgrpstart
\figsetgrpnum{4.28}
\figsetplot{smc27_gf.pdf}
\label{fig:smc27}
\figsetgrpnote{Plots for SMC27, a field with two PDs (labeled as P2). Panel (a) shows a polarization map overlapped with a {\it Spitzer}/MIPS image at $160~\mu$m \citep{bib:gordon11}; panel (b) shows a plot of $\theta$
versus magnitude; panel (c) shows a $\theta$ histogram, the green and gold lines represent the Gaussian fits for the first
and second PDs, respectively; and panel (d)
shows a polarization intensity histogram, the vertical lines on this plot represents the median value for the first (solid line) and second PD
(solid-dotted line).}
\figsetgrpend

\figsetend

\subsection{Fields with No PD (P0)}

	The fields SMC 05, 14, 19, and 25 did not show any PD, therefore we did not fit any Gaussian to them. Figure \ref{fig:smc14} shows
the example of SMC14: all the stars in the field were considered to construct the $\theta$ and $P$ histograms. This field shows a
completely random distribution. Hence, we are not able to define any PD for the polarization. In the case of SMC05 (Figure \ref{fig:smc05}c), there is
possibly a PD; however, the number of stars is not high enough to characterize it. These fields are represented by gray squares in Figure \ref{fig:fields}.

\begin{figure*}[!htb]\centering
\figurenum{4.1}
\includegraphics[width=1.0\textwidth]{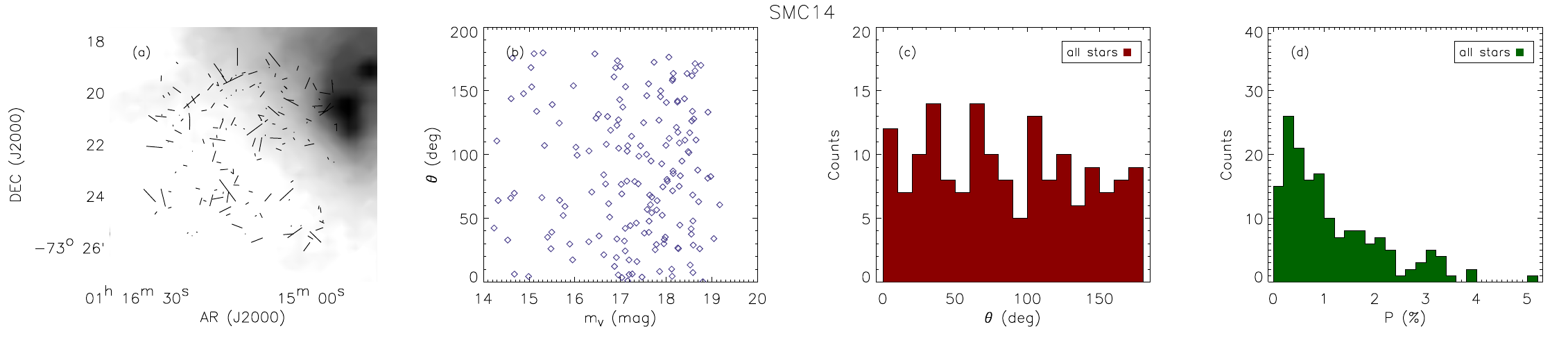}
\caption{Plots for SMC14, an example of a field with no PD (labeled as P0). Panel (a) shows a polarization map overlapped with a {\it Spitzer}/MIPS image at $160~\mu$m \citep{bib:gordon11}; panel (b) shows a plot of $\theta$
versus magnitude; panel (c) shows a $\theta$ histogram; and panel (d)
shows a polarization intensity histogram.}
\label{fig:smc14}
\end{figure*}

\begin{figure*}[!htb]\centering
\figurenum{4.2}
\includegraphics[width=1.0\textwidth]{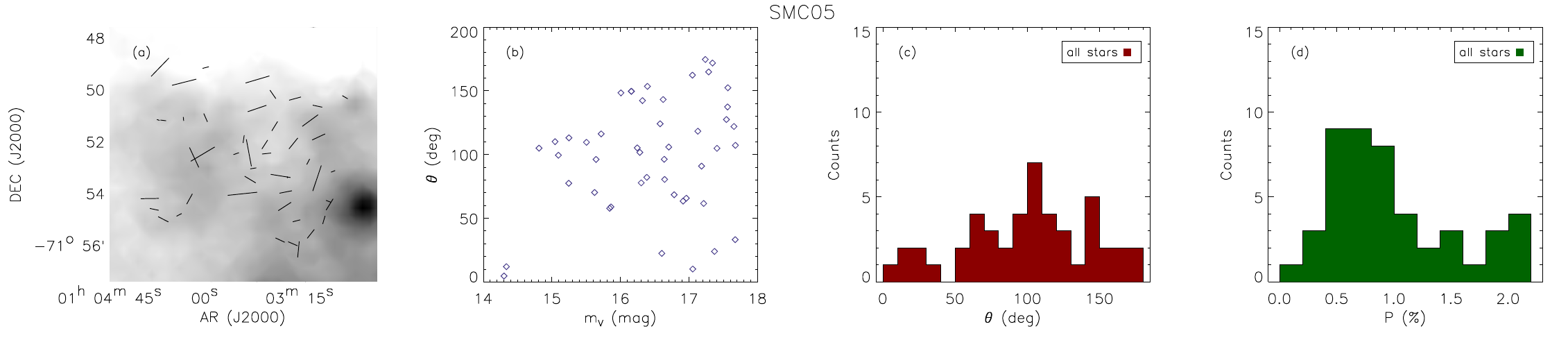}
\caption{Plots for SMC05, an example of a field with no PD (labeled as P0). Panel (a) shows a polarization map overlapped with a {\it Spitzer}/MIPS image at $160~\mu$m \citep{bib:gordon11}; panel (b) shows a plot of $\theta$
versus magnitude; panel (c) shows a $\theta$ histogram; and panel (d)
shows a polarization intensity histogram.}
\label{fig:smc05}
\end{figure*}

\subsection{Fields with One PD (P1)}

	The $\theta$ histogram for fields SMC 03, 06, 09, 11, 15, 18, 21, and 28 showed one PD. Consequently one Gaussian was fitted for each of these. Some
fields show a clear PD, an example is SMC28 (Figure \ref{fig:smc28}c). In others, SMC09 for instance (Figure \ref{fig:smc09}c), it is almost possible to
see a second PD; however, the number of stars is once again not sufficient for the Gaussian fitting. Finally, some fields displayed a large
dispersion, an example is SMC03 (Figure \ref{fig:smc03}c), but not large enough to qualify the distribution's nature as random. The red squares in Figure \ref{fig:fields} represent these fields.

\begin{figure*}[!htb]\centering
\figurenum{4.3}
\includegraphics[width=1.0\textwidth]{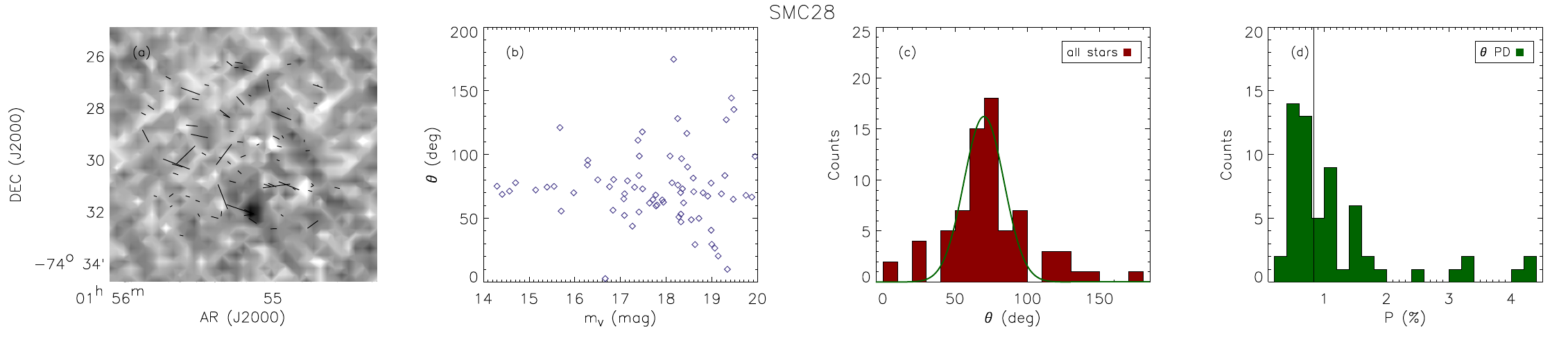}
\caption{Plots for SMC28, an example of a field with one PD (labeled as P1). Panel (a) shows a polarization map overlapped with a {\it Spitzer}/MIPS image at $160~\mu$m \citep{bib:gordon11}; panel (b) shows a plot of $\theta$
versus magnitude; panel (c) shows a $\theta$ histogram, the green line represents the Gaussian fit; and panel (d)
shows a polarization intensity histogram, the vertical line on this plot represents the median value.}
\label{fig:smc28}
\end{figure*}

\begin{figure*}[!htb]\centering
\figurenum{4.4}
\includegraphics[width=1.0\textwidth]{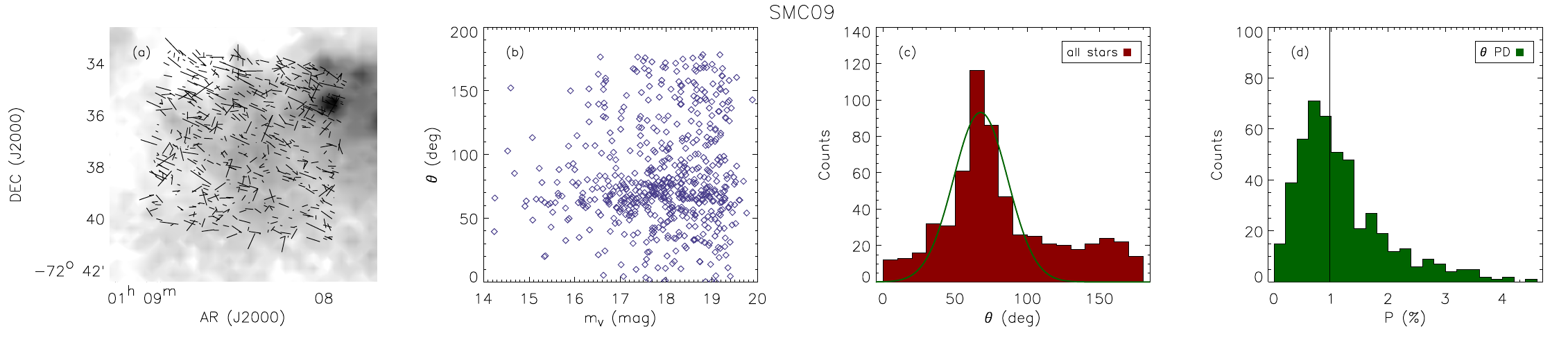}
\caption{Plots for SMC09, an example of a field with one PD (labeled as P1). Panel (a) shows a polarization map overlapped with a {\it Spitzer}/MIPS image at $160~\mu$m \citep{bib:gordon11}; panel (b) shows a plot of $\theta$
versus magnitude; panel (c) shows a $\theta$ histogram, the green line represents the Gaussian fit; and panel (d)
shows a polarization intensity histogram, the vertical line on this plot represents the median value.}
\label{fig:smc09}
\end{figure*}

\begin{figure*}[!htb]\centering
\figurenum{4.5}
\includegraphics[width=1.0\textwidth]{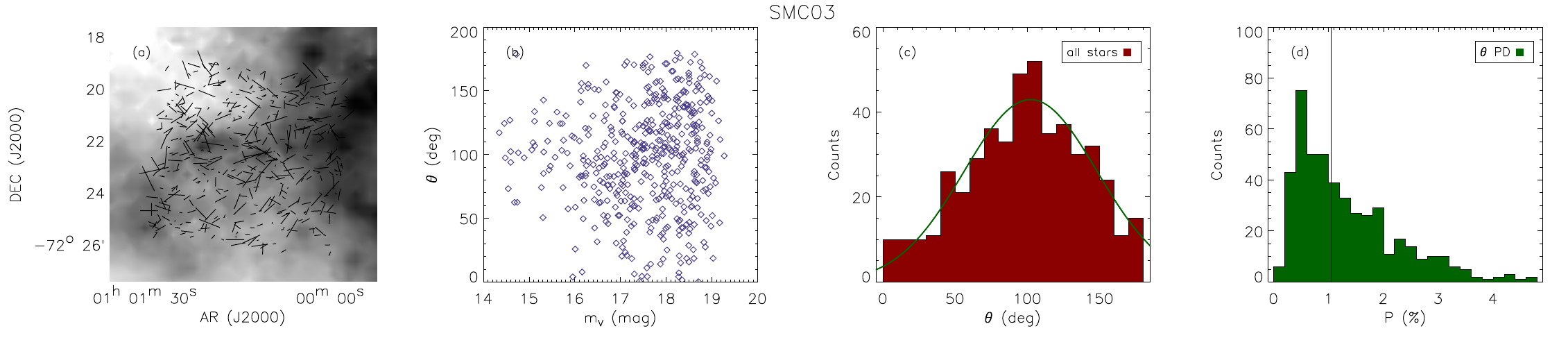}
\caption{Plots for SMC03, an example of a field with one PD (labeled as P1). Panel (a) shows a polarization map overlapped with a {\it Spitzer}/MIPS image at $160~\mu$m \citep{bib:gordon11}; panel (b) shows a plot of $\theta$
versus magnitude; panel (c) shows a $\theta$ histogram, the green line represents the Gaussian fit; and panel (d)
shows a polarization intensity histogram, the vertical line on this plot represents the median value.}
\label{fig:smc03}
\end{figure*}

\subsection{Fields with Two PDs (P2)}

	Two PDs for fields SMC 01, 02, 04, 07, 08, 10, 22, and 27 were observed in the $\theta$ histogram. Similarly, two Gaussians were fitted to
reproduce the two PDs.
Figure \ref{fig:smc01}c shows the $\theta$ histogram for SMC01, which, along with other fields, a possible third PD can be discerned.
Gaussians were just fitted to those PDs that are significant enough relative to the random background and distinguishable from the neighboring PD.
In the other hand, some fields display two clear PDs, an example is SMC10 (Figure
\ref{fig:smc10}c). The magenta squares in Figure \ref{fig:fields} represent these fields.

\begin{figure*}[!htb]\centering
\figurenum{4.6}
\includegraphics[width=1.0\textwidth]{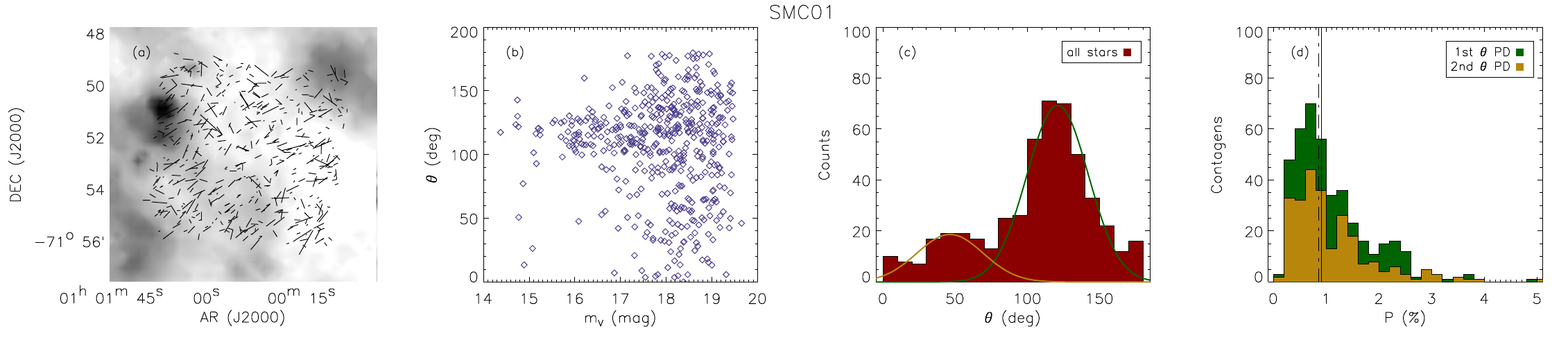}
\caption{Plots for SMC01, an example of a field with two PDs (labeled as P2). Panel (a) shows a polarization map overlapped with a {\it Spitzer}/MIPS image at $160~\mu$m \citep{bib:gordon11}; panel (b) shows a plot of $\theta$
versus magnitude; panel (c) shows a $\theta$ histogram, the green and gold lines represent the Gaussian fits for the first
and second PDs, respectively; and panel (d)
shows a polarization intensity histogram, the vertical lines on this plot represents the median value for the first (solid line) and second PD
(solid-dotted line).}
\label{fig:smc01}
\end{figure*}

\begin{figure*}[!htb]\centering
\figurenum{4.7}
\includegraphics[width=1.0\textwidth]{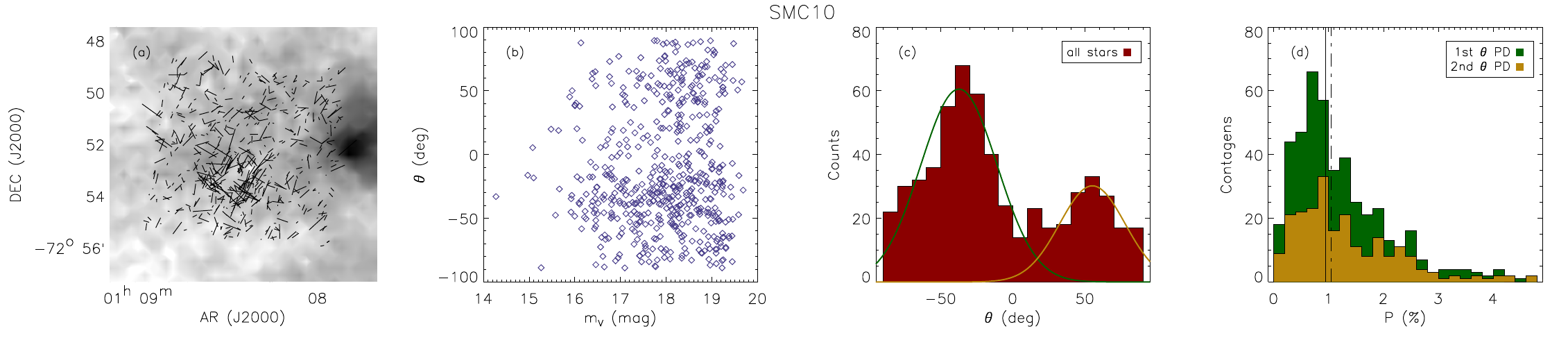}
\caption{Plots for SMC10, an example of a field with two PDs (labeled as P2). Panel (a) shows a polarization map overlapped with a {\it Spitzer}/MIPS image at $160~\mu$m \citep{bib:gordon11}; panel (b) shows a plot of $\theta$
versus magnitude; panel (c) shows a $\theta$ histogram, the green and gold lines represent the Gaussian fits for the first
and second PDs, respectively; and panel (d)
shows a polarization intensity histogram, the vertical lines on this plot represents the median value for the first (solid line) and second PD
(solid-dotted line).}
\label{fig:smc10}
\end{figure*}

\subsection{Fields Filtered by Magnitudes (PFM)}

	The $\theta$ histogram for fields SMC 12, 13, 20, and 23 are the result of a PD and a random background being superposed. In these fields, the peak position
was well fitted, but the standard deviation was not. To improve the fits, we
made plots of $\theta$ vs $m_V$, to check whether it is possible to separate the random background from the PD. Figure \ref{fig:smc12} shows, for
SMC12, the histograms for $\theta$ and $P$, as well as the plot for $\theta$ vs $m_V$. For this specific case, the random background is
due to stars with $m_V>17.5~$mag. Hence, we selected the stars with magnitudes smaller than this value to perform the Gaussian fit. This cut in magnitude can be
justified, taking into account that magnitudes are distance indicators, therefore the objects forming the random distribution may constitute a further
population. In order to determine the optimum value of $m_V$, trying to separate the random background, we used plots of the dispersion of $\theta$ in function of
$m_V$. More details about this procedure can be seen in the Appendix \ref{sec:appendixc}. These fields are represented by steel blue squares in Figure \ref{fig:fields}.

\begin{figure*}[!htb]\centering
\figurenum{4.8}
\includegraphics[width=1.0\textwidth]{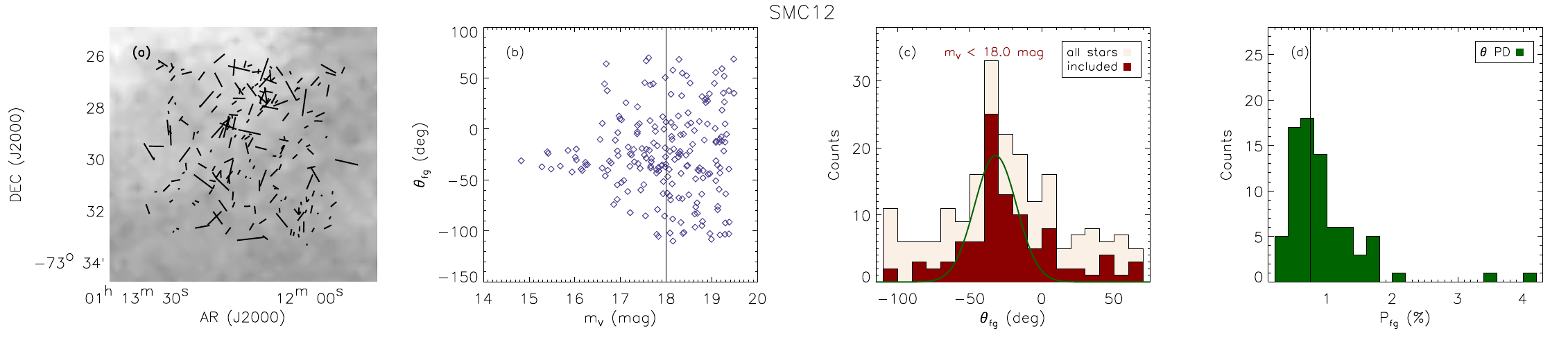}
\caption{Plots for SMC12, an example of a field for which we had to filter the stars by magnitudes to fit the polarization PD (labeled as PFM). Panel
(a) shows a polarization map overlapped with a {\it Spitzer}/MIPS image at $160~\mu$m \citep{bib:gordon11}; panel (b) shows a plot of $\theta$ versus magnitude, the vertical line demarcates the magnitude cut; panel (c)
shows a $\theta$ histogram, the green line represents the Gaussian fit; and panel (d)
shows a polarization intensity histogram, the vertical line on this plot represents the median value.}
\label{fig:smc12}
\end{figure*}

\subsection{Fields with Stars Excluded (PSE)}

	For the remaining fields, SMC 16, 17, 24, and 26, a simple Gaussian fit does not converge if all objects are considered. In the case of SMC24,
there are two PDs in the $\theta$ vs $m_V$ plot (Figure \ref{fig:smc24}b). One of them is composed of few stars, such that two Gaussian fits were
not possible. Hence, we decided to exclude these stars. For the other three fields, we could not observe any different
behavior along the magnitude range, an example is SMC16 (Figure \ref{fig:smc16}b). In order to be able to perform the Gaussian fits, we
arbitrarily excluded the stars that were preventing
the fit convergence. The spring green squares in Figure \ref{fig:fields} represent these fields.

\begin{figure*}[!htb]\centering
\figurenum{4.9}
\includegraphics[width=1.0\textwidth]{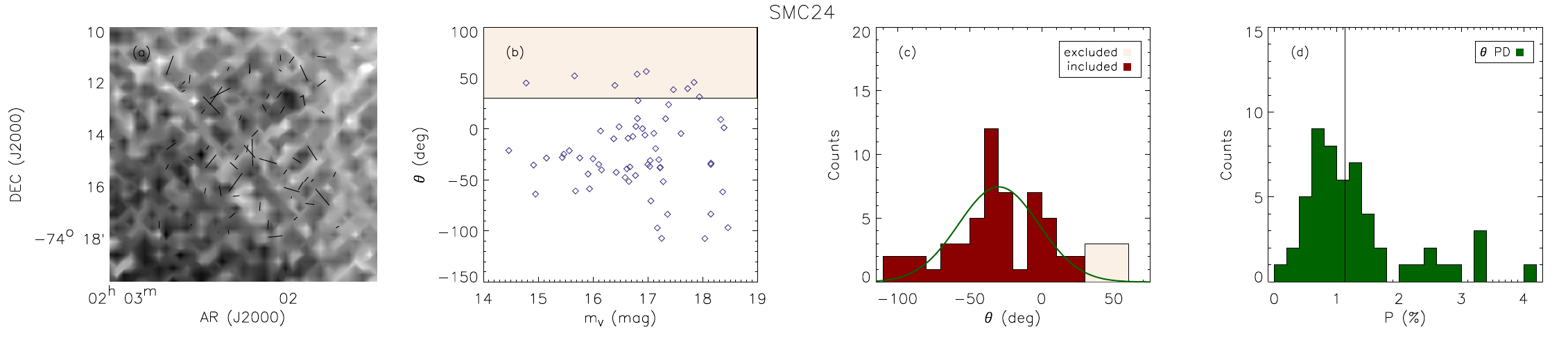}
\caption{Plots for SMC24, an example of a field for which we had to exclude some stars to fit the polarization PD (labeled as PSE). Panel (a) shows a
polarization map overlapped with a {\it Spitzer}/MIPS image at $160~\mu$m \citep{bib:gordon11}; panel (b) shows a plot of $\theta$ versus magnitude, the stars in the light pink area were not considered for the
Gaussian fit; panel (c) shows a $\theta$ histogram, the green line represents the Gaussian fit; and panel (d)
shows a polarization intensity histogram, the vertical line on this plot represents the median value.}
\label{fig:smc24}
\end{figure*}

\begin{figure*}[!htb]\centering
\figurenum{4.10}
\includegraphics[width=1.0\textwidth]{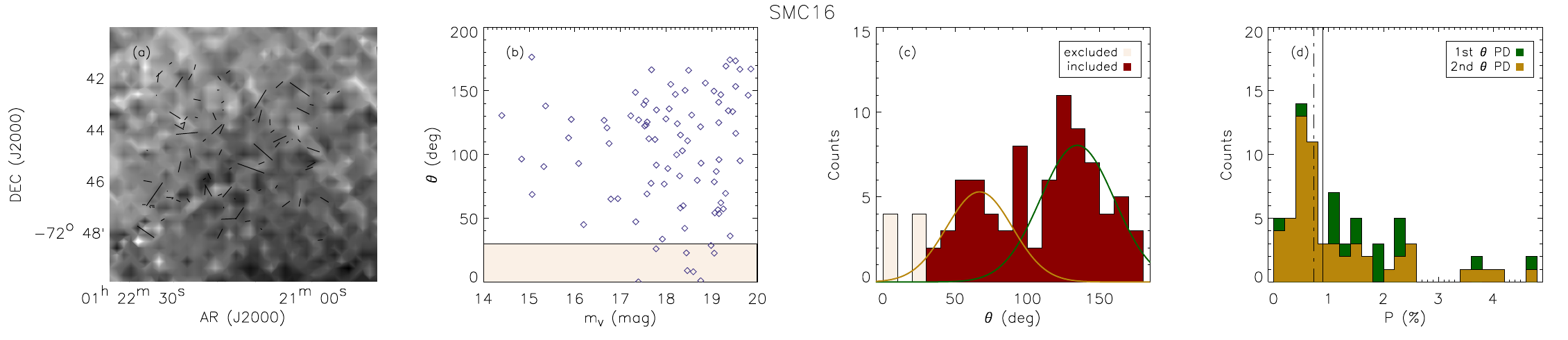}
\caption{Plots for SMC16, an example of a field for which we had to exclude some stars to fit the polarization PD (labeled as PSE). Panel (a) shows a
polarization map overlapped with a {\it Spitzer}/MIPS image at $160~\mu$m \citep{bib:gordon11}; panel (b) shows a plot of $\theta$ versus magnitude, the stars in the light pink area were not considered for the
Gaussian fit; panel (c) shows a $\theta$ histogram, the green and gold lines represent the Gaussian fits for the first
and second PDs, respectively; and panel (d)
shows a polarization intensity histogram, the vertical lines on this plot represents the median value for the first (solid line) and second PD
(solid-dotted line).}
\label{fig:smc16}
\end{figure*}


\section{Magnetic Field Geometry}\label{sec:Bgeometry}

\begin{figure}[!htb]\centering
\figurenum{5}
\includegraphics[width=1.\columnwidth]{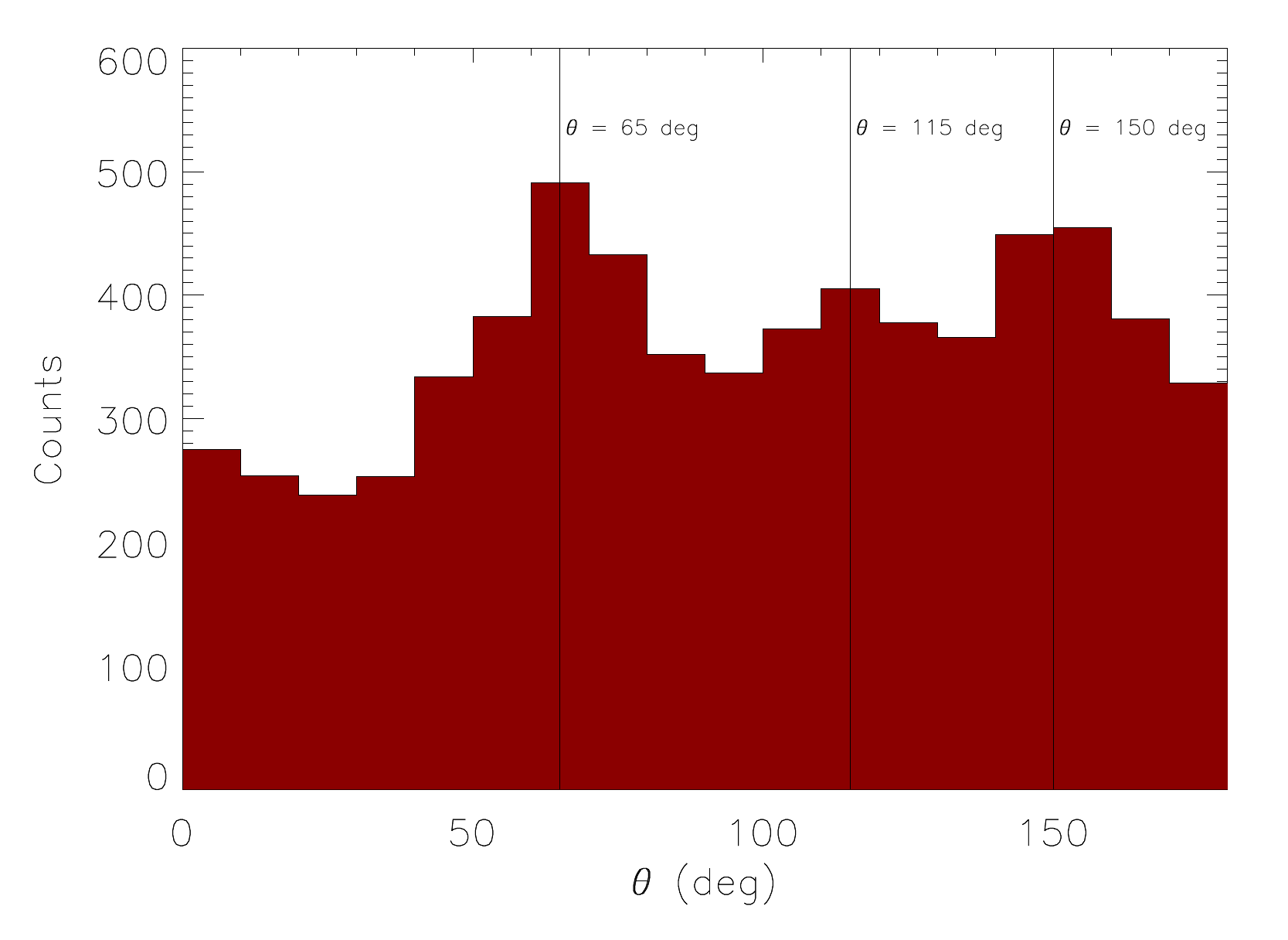}
\caption{$\theta$ histogram of the data corrected by foreground from this work and with $P/\sigma_P>3$. The vertical lines demarcate the trends.}
\label{fig:PAhist}
\end{figure}

	In this Section we discuss the magnetic field geometry of the SMC based on polarization maps.
Initially, a preliminary check was made by constructing a $\theta$ histogram in order to check the
general behavior of the data. We used the data that was foreground-corrected with the foreground estimate from this work (see Section \ref{sec:foreground}) and
considered objects with intrinsic polarization: $P/\sigma_P>3$. Figure \ref{fig:PAhist} shows the $\theta$ histogram.

\begin{figure*}[!htb]\centering
\figurenum{6}
\includegraphics[width=.9\textwidth]{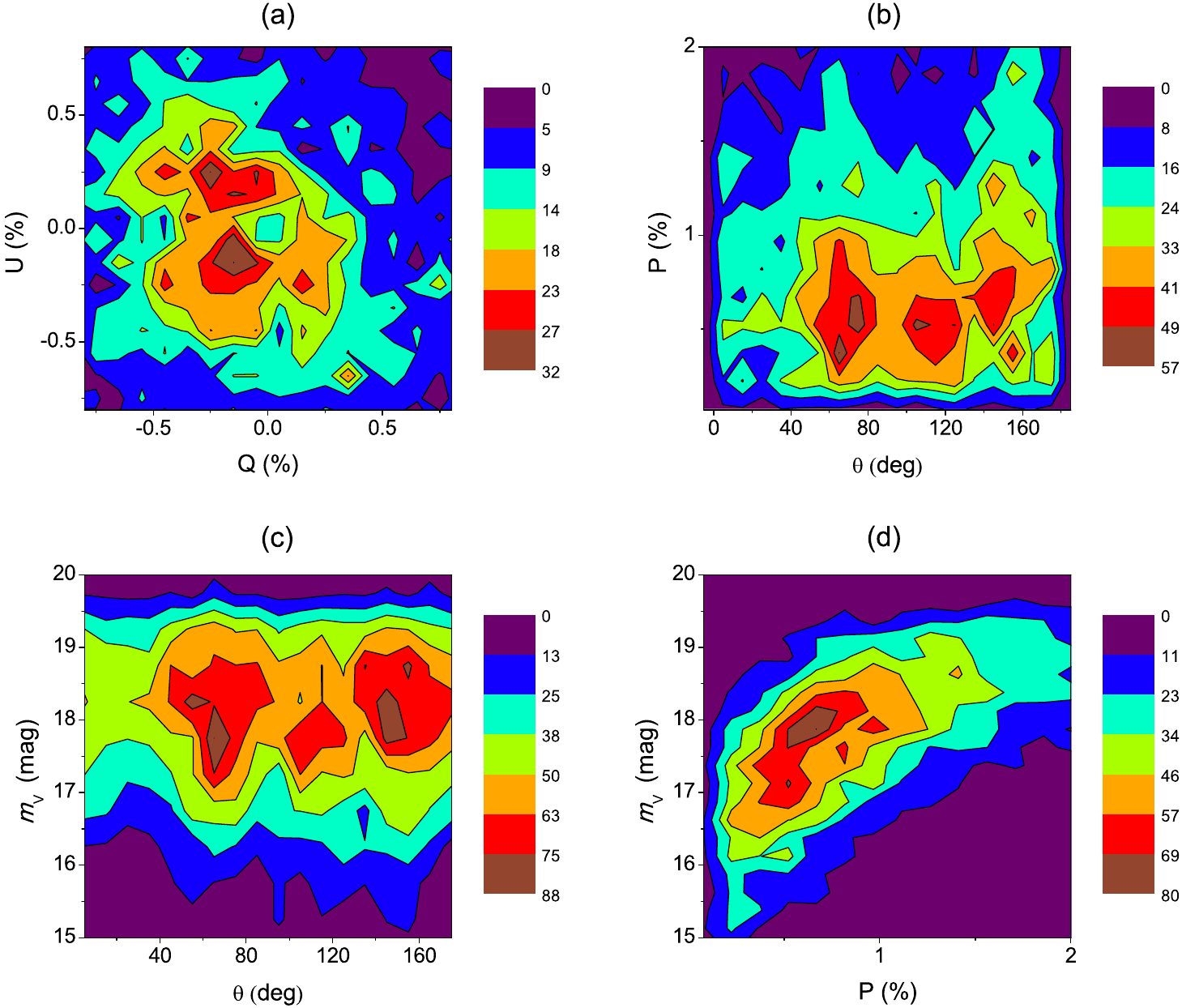}
\caption{Density plots of the SMC intrinsic polarization. Panel (a) shows a plot of $U$ vs. $Q$, panel (b) of $P$ vs. $\theta$,
panel (c) of $m_V$ vs. $\theta$,
and panel (d) of $m_V$ vs. $P$. The color bars
indicate the number of objects.}
\label{fig:densityInt}
\end{figure*}

	The histogram in Figure \ref{fig:PAhist} reveals three major trends in the sample. The most prominent is featured at ($65^{\circ}\pm10^{\circ}$) (trend I). The latter two trends were identified at
($115^{\circ}\pm10^{\circ}$) (trend II) and ($150^{\circ}\pm10^{\circ}$) (trend III). The errors for each trend were estimated as the bin size considered
for the histogram. To assess whether these trends are either separated in distance or if any kind of segregation is present,
density plots ($Q$ vs. $U$, $\theta$ vs. $P$, $\theta$ vs. $m_V$, and $P$ vs. $m_V$) were constructed, as can be seen in 
Figure \ref{fig:densityInt}. These trends were further reinforced in the plots of $Q$ vs. $U$, $\theta$ vs. $P$, and $\theta$ vs. $m_V$. We do not observe any
strong segregation; however, trend II is concentrated at brighter stars (in the range $\sim 17.2$ to $\sim 18.25$~mag) and smaller polarization intensities (in the range $\sim 0.25\%$ to
$\sim 0.75\%$). This indicates that trend II could be associated with the part of the bimodal structure that lies closer to us in distance. Trend I encompasses a wide
range in polarization (from $\sim 0.2\%$ to $\sim 0.9\%$) and magnitude (from $\sim 17.1$ to $\sim 18.9$~mag), while trend III is shifted in the
vertical direction in both polarization (from $\sim 0.4\%$ to $\sim 0.8\%$) and magnitude (from $\sim 17.5$ to $\sim 19$~mag), but
concentrated in a smaller range of polarization intensities. In the $P$ vs. $m_V$ plot no segregation is observed, just the normal behavior of polarization increasing with magnitude.

\begin{figure*}[!htb]\centering
\figurenum{7}
\includegraphics[width=.9\textwidth]{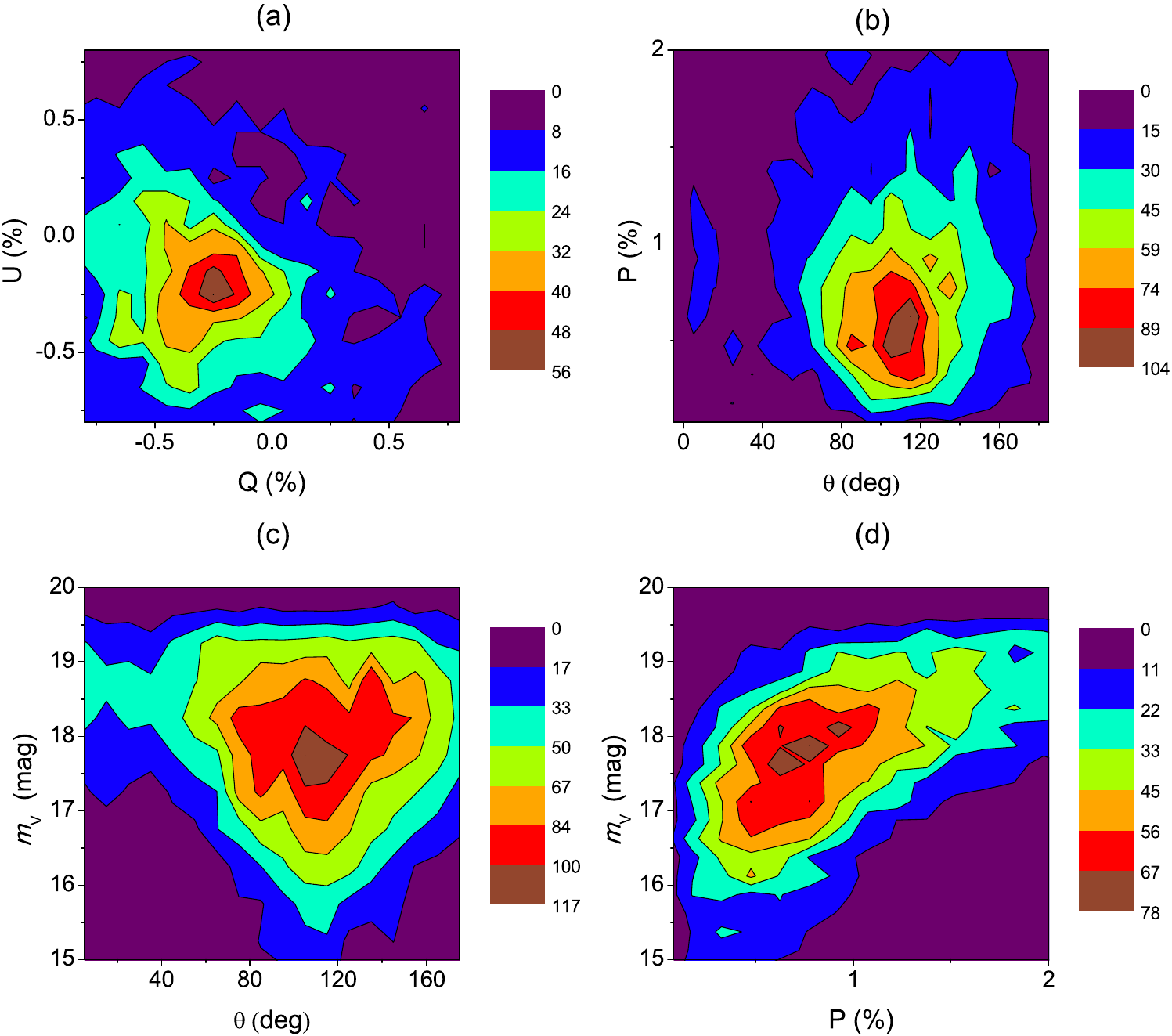}
\caption{Density plots of the SMC observed polarization. Panel (a) shows a plot of $U$ vs. $Q$, panel (b) of $P$ vs. $\theta$, panel
(c) of $m_V$ vs. $\theta$,
and panel (d) of $m_V$ vs. $P$. The color bars
indicate the number of objects.}
\label{fig:densityObs}
\end{figure*}

	In order to check the impact of the foreground subtraction on the data, we made density plots of the same quantities as before, but
using the SMC observed polarization (Figure \ref{fig:densityObs}). For the observed data, just trend II is visible, which demonstrates how much the
foreground subtraction changes the geometry observed, raising two other trends that were masked by the foreground contribution.

	Based on the discussion above, we separated the PDs obtained in the previous section in to three groups:

\begin{enumerate}
\item $40^{\circ}$ $\leq \theta_{\text{PD}} \leq 90^{\circ}$ (trend I);
\item $90^{\circ}$ $< \theta_{\text{PD}} \leq 132^{\circ}.5$ (trend II);
\item $132^{\circ}.5$ $< \theta_{\text{PD}} \leq 185^{\circ}$ (trend III). 
\end{enumerate}

	The upper and lower limits for each group
were chosen to be the position of its trend plus/minus half
distance to the neighbor trend. Trend I has no neighbor in the left side, therefore its lower limit was defined as the position of trend I minus 
half distance between trend I and II. Trend III has no neighbor in the right side, therefore its upper limit was defined as a value that includes
all the PDs lying in its right side.

\begin{figure*}[!htb]\centering
\figurenum{8}
\includegraphics[width=.9\textwidth]{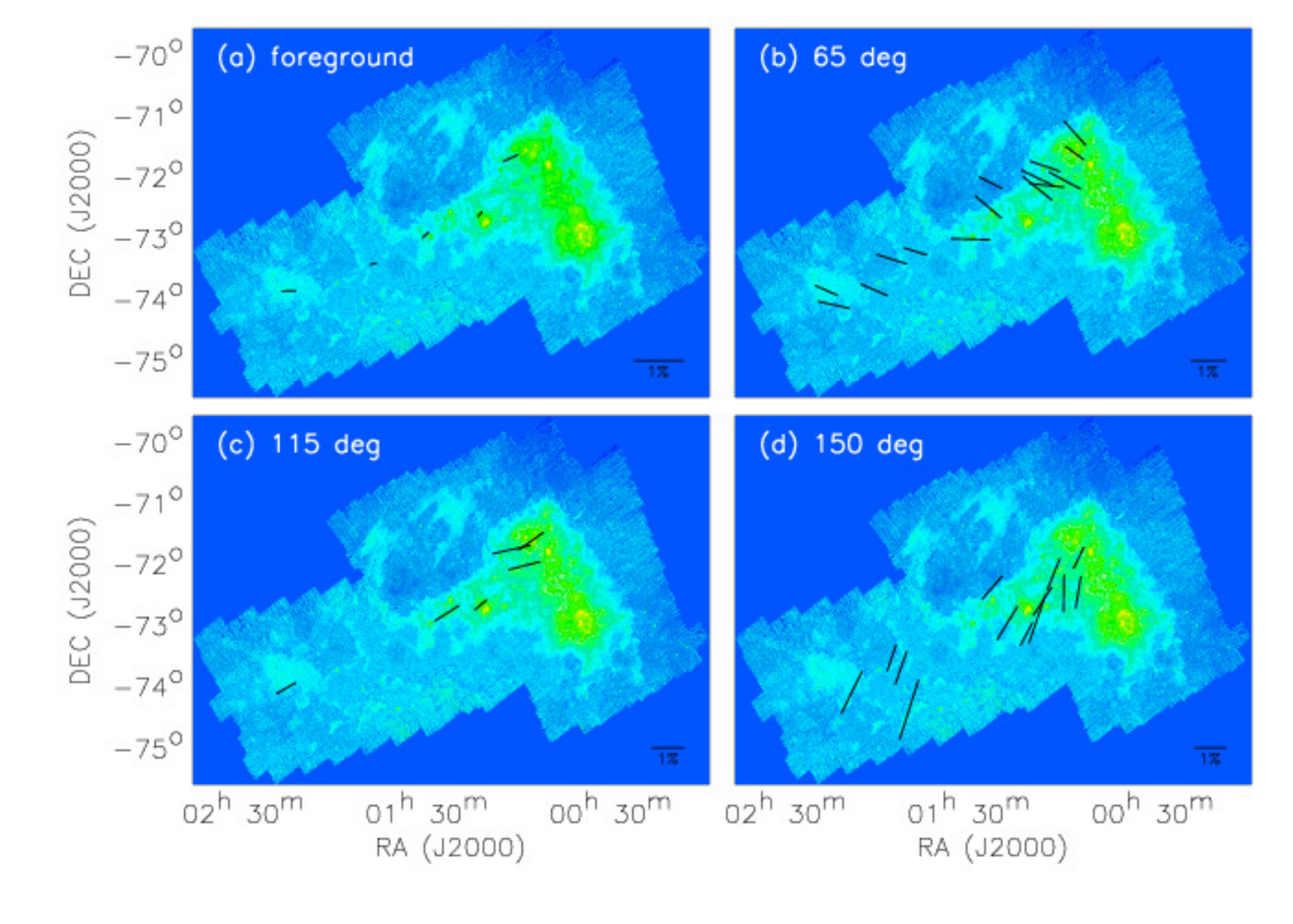}
\caption{Polarization maps. Panel (a) shows a map of the foreground polarization from this work, panel (b) shows a map of the PDs in the range
[$40^{\circ}.0-90^{\circ}.0$], panel (c) shows a map of the PDs in the range [$90^{\circ}.0-132^{\circ}.5$], and panel (d) shows a map of the PDs in the range
[$132^{\circ}.5-185^{\circ}.0$]. The vectors are overlapped with a {\it Spitzer}/MIPS image at $160~\mu$m \citep{bib:gordon11}. The Magellanic Bridge stars in the
lower left region of the maps and extends up to the LMC location.}
\label{fig:maps}
\end{figure*}

	Each group contains one of the trends, thus all the PDs were classified as belonging to a trend. Figure \ref{fig:maps} shows the maps obtained
for each group and the foreground map as determined by this work. The foreground map (Figure \ref{fig:maps}a) shows that the Galactic foreground is highly aligned with the Bridge
direction at a PA of $115^{\circ}.4$. Figure \ref{fig:maps}b shows that the fields in the NE region present an ordered magnetic
field roughly aligned with the Bar at a PA of $\sim45^{\circ}$. Figure \ref{fig:maps}c also shows an alignment with the Bridge direction;
however, it is the trend containing the least number
of vectors. Figure \ref{fig:maps}d shows vectors separated by about $90^{\circ}$ from the ones in Figure \ref{fig:maps}b. Overall the polarization maps do
not show a very demarcated structure, but rather a quite complex geometry.

\subsection{Magnetic Field Origin}

	The origin of the SMC's large scale magnetic field was discussed by \cite{bib:mao08}. This study concluded that a cosmic-ray driven dynamo can explain the existence of a large scale magnetic
field at the SMC in terms of time scale arguments. Nonetheless, it had difficulties in explaining the geometry observed, because of the
unidirectionality of the magnetic field and no change of sign for the Faraday rotation measures (RM). Considering that the cosmic-ray driven dynamo
is the mechanism that generated the large scale magnetic field at the SMC, initially this field had
mostly an azimuthal configuration, which is the axisymmetric dynamo mode $m=0$. Tidal interactions lead to the excitation of the bisymmetric mode $m=1$,
when the axisymmetric mode is already at work \citep{bib:moss95}. The tidal interactions also induce nonaxisymmetric velocities in the interacting
galaxies disks and may lead to the damping of the $m=0$ mode, leaving just the bisymmetric magnetic field \citep{bib:vogler01}.

	\cite{bib:mao08} explained that in the scenario above, the fact that the SMC's magnetic field is predominantly lying on its disk, can be explained by the fact that the azimuthal magnetic field produced by a
dynamo mostly lies in the disk of the galaxy. The inclusion of tidal forces between the SMC and LMC would explain the alignment with the Magellanic
Bridge. Finally, if the SMC possesses a bisymmetric magnetic field, we would observe a periodic double change of the RM sign with respect to the
azimuthal angle. If the
magnetic field is represented by a superposition of $m=0$ and $m=1$ modes, even more sign changes would be expected. Considering that the magnetic field lines do
close, but the locations with field lines pointing toward us are outside the SMC's body, this would explain why it is observed just negative RMs. The regions that should
display positive RMs may have a low emission measure of ionized gas, therefore RM is zero in this locations, since to observe non-zero RM the average
electron density in the line of sight should be non-zero. Our data do not exclude the physical explanation given by \cite{bib:mao08}. On the
contrary, the unidirectionality for the magnetic field
is no longer a problem. 

	Trend I is widely spread in magnitudes and polarizations, which may indicate its correlation with a global pattern of the SMC.
\cite{bib:stanimirovic04} obtained that the PA for the major kinematic axis of the SMC is around $50^{\circ}$, which is $15^{\circ}$ in difference
from trend I. This direction is also roughly the PA of the SMC's Bar, positioned at $\sim45^{\circ}$ \citep{bib:bergh07}. This suggests that in the Bar region the
magnetic field may be coupled to the gas and therefore the field lines are roughly parallel to the Bar direction due to the flux freezing condition; however, we can not explain why the field lines in the SMC's
Wing and in the Magellanic Bridge display also an alignment with this direction. Our understanding is that the initial $m=0$ dynamo mode may have been damped by the
nonaxisymmetric velocities excited by the tidal interactions. Therefore we do not observe the typical behavior of an azimuthal field in the polarization
vectors. Nonetheless, the bisymmetric field was left and possibly higher order dynamo components. The current geometry for the SMC's magnetic field is
probably the product of an active interaction of the SMC--LMC--MW system summed to the influence of star formation, supernova explosions, and other
processes that can inject energy to the ISM, controlling its dynamics. \cite{bib:burkhart10} suggests that the HI in the SMC is super-Alfv\'enic,
which also explains the rather disordered configuration observed for the large scale magnetic field.

	As mentioned before, our data are
concentrated at the NE and Wing sections of the SMC and at a part of the Magellanic Bridge. In these regions a bimodality in the distance is known to
exist from Cepheid distances \citep{bib:mfv86,bib:nidever13}. Similarly, two velocity components are observed in HI studies
\citep{bib:mathewson84,bib:stanimirovic04}. Figure \ref{fig:densityInt} indicates that trend II could be related to the
component located closer to us and that trend III could be linked to the most distant component. The difference in magnitudes between trends II and
III is about $0.5$~mag. Translating into relative distances: $d_{III} = 1.26d_{II}$, neglecting internal extinction. Hence, if trend II belongs to the closest
component, located at $55$~kpc as observed from Cepheids in the eastern region \citep{bib:nidever13}, trend III may be located at $69$~kpc.
Considering the photometric error of our catalog of $0.13$~mag, the distance for trend III obtained by this rough estimate is compatible with the
distance obtained by Cepheids for the furthest component, which is $67$~kpc \citep{bib:nidever13}. The tidal interactions between SMC and LMC
are likely to explain the stretching of the magnetic field lines toward the Magellanic Bridge direction. Nonetheless, this effect was important just in the component closer to us in
distance, which is also closer to the LMC. The creation of bridges and the magnetic field alignment with respect to the bridge was already observed
in numerical studies. \cite{bib:kotarba11} simulated the interaction of three disk galaxies up to the point where they all merge. Their simulation
shows that the magnetic field of the interacting galaxies strongly changes with time according to their interaction. Nonetheless, the comparison of the SMC--LMC--MW system with their
results is not straightforward, because the properties of the galaxies are different.

	The coincidence between the directions of trend II and the Galactic foreground is not easy to explain. The magnetic field at the Galactic halo
is not expected to be high and indeed the polarization measurements are rather low in that region ($P_{for} \lesssim 0.5\%$). A possible speculation is that the MW
halo also feels the tidal forces
by the SMC and LMC, therefore its magnetic field is also stretched in the same direction. The simulation by \cite{bib:kotarba11} shows that when the
galaxies are about $50$~kpc apart their magnetic fields align in the outskirts of the approaching galaxies (Figure 15 in their paper). The masses,
magnetic fields strengths, and 3D distribution of the SMC--LMC--MW system is not alike their system. Nonetheless, we speculate that the coincidence of
trend II with the Galactic foreground can be explained by the system interaction.

	Yet, a question can be raised regarding the genuinity of trend II: is it real or a vestige of a bad
foreground removal? This question is difficult to address since the mentioned direction is that of the Galactic foreground. If the foreground was underestimated, the trend could be a remnant of the
foreground itself; however, the median polarizations for trend II ($0.46\%~\leq~P_{II}~\leq~1.2\%$) range two to eight times above the estimated
foreground ($0.06\%~\leq~P_{for}~\leq~0.47\%$). Therefore a foreground underestimation can be justifiably ruled out. Another
possibility is that faint stars from the MW are included in the catalog, causing the foreground trend to persist. To fully rule this out, the distance to those stars or another distance indicator such
as the $E(B-V)$ color excess is necessary to separate SMC and MW members. The upcoming GAIA mission will prove to be a good tool for SMC--MW member separation
due to the parallaxes that will be measured in the MW halo. The accuracy expected for the fainter stars ($m_V = 20$~mag) is to be as good as
$1$~mas, about $10\%$ for stars at $10$~kpc. The stars that form trend II have magnitudes from $17.2$ to $18.3$~mag, thus the GAIA accuracy
may be good enough to define whether these stars belong to the MW's halo or the SMC.

	This work brings a new understanding of the SMC's magnetic field. Nevertheless, the SOUTH POL project \citep{bib:southpol} will measure the
polarization of objects in the whole southern sky (south of $-15^{\circ}$ initially), which will increase even more the
polarization sample toward the SMC. These new data will help to get a more complete picture of the SMC magnetic field structure, since it will
measure polarizations in regions not covered by this work. Using the
photometric catalog of SMC members of \cite{bib:massey02}, to get the number of stars per magnitude range, we expect that polarizations will be
measured for around 7500 bright stars ($m_V \lesssim 15$~mag) with accuracy up to $0.1\%$ and around 68,000 stars
($m_V \lesssim 17$~mag) with accuracy up to $0.3\%$. Naturally, fainter stars will de detected, but with accuracies that may or may not
be appropriate for ISM studies; for instance, stars with $m_V \sim 18$~mag will be measured with about $1\%$ of accuracy. Moreover, SOUTH POL will
measure additional foreground objects toward the SMC, which should have GAIA distance estimates.


\section{Alignment between the SMC Magnetic Field and the Magellanic Bridge}\label{sec:alignment}

	To further address the question regarding the alignment between the SMC polarization and Magellanic Bridge direction, we used cumulative frequency
distribution (CFD) analysis. Firstly, we calculated the angle between the SMC and LMC
centers: $\theta_{0M} = 115^{\circ}.4$, which is in the same direction as the Magellanic Bridge. Similarly to \cite{bib:schmidt76}, we defined $\theta_M$
as the angles between $\theta_{0M}$ and the
polarization angles of the stars. The following coordinates were used for the centers of the SMC and LMC respectively:
R.A.(J2000)~=~$00^h52^m38.0^s$, decl.(J2000)~=~$-72^{\circ}48'01''$ and R.A.(J2000)~=~$05^h23^m34.6^s$,
decl.(J2000)~=~$-69^{\circ}45'22''$. 

\begin{figure*}[!htb]\centering
\figurenum{9}
\includegraphics[width=1.\textwidth]{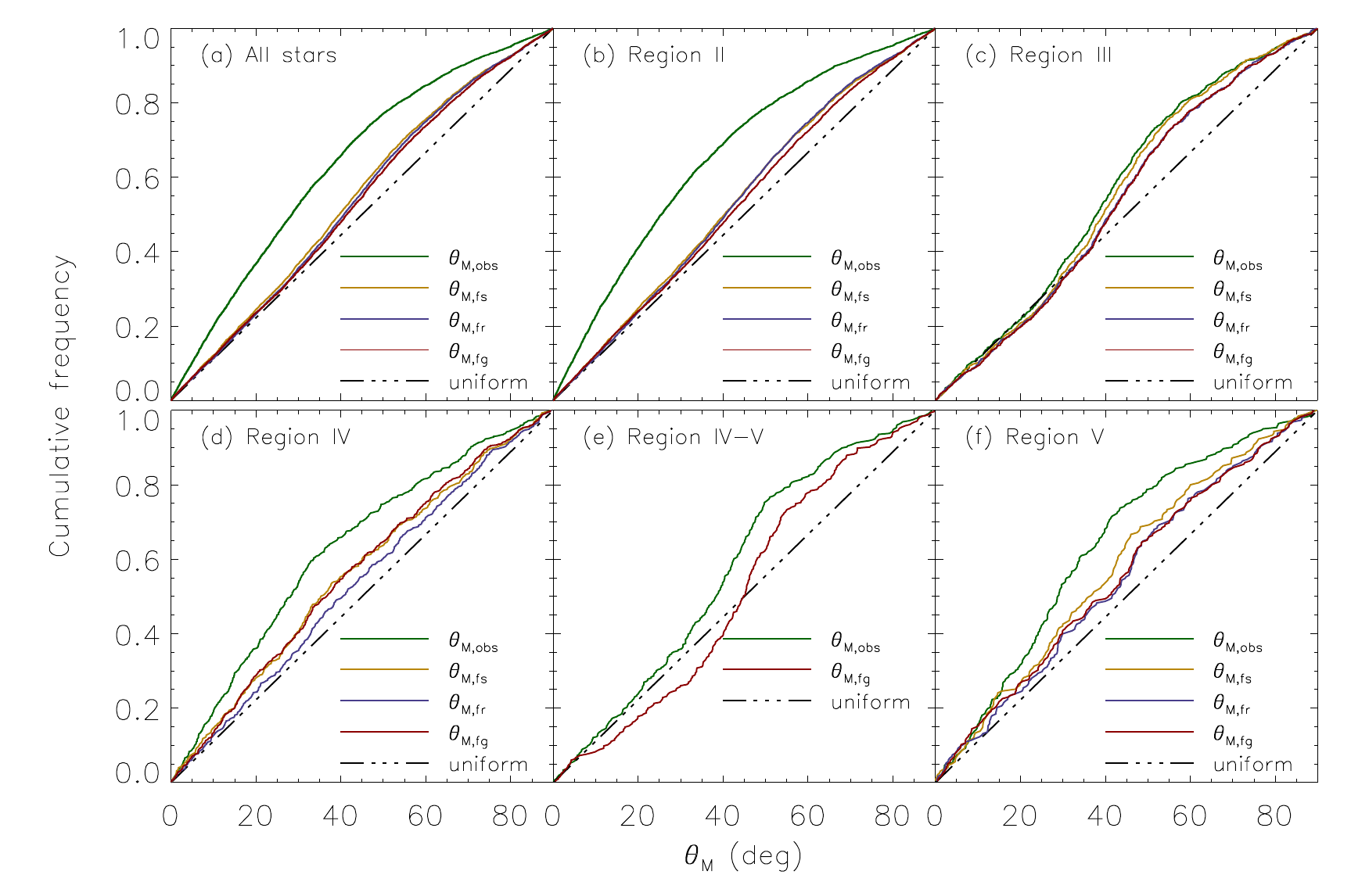}
\caption{Cumulative frequency distribution for $\theta_M$, the angle between the polarization angles and the SMC--LMC direction. The plot on the upper
left, panel (a), shows the CFD for all stars; on the upper middle, panel (b), for region II; on the upper right, panel (c), for region III; on the bottom
left, panel (d), for region IV; on the bottom middle, panel (e), for region IV--V; and on the bottom right, panel (f), for region V. $\theta_{M,obs}$
corresponds to the observed polarization, $\theta_{M,fs}$ corresponds to polarization corrected by \cite{bib:schmidt76} foreground, $\theta_{M,fr}$
corresponds to polarization corrected by \cite{bib:rodrigues97} foreground and $\theta_{M,fg}$ corresponds to polarization corrected by foreground
from this work. The dot-dashed lines correspond to a uniform distribution.}
\label{fig:cfd}
\end{figure*}

	Considering stars with $P/\sigma_P > 3$ and $m_V > 14.2$~mag, CFDs were constructed. Table \ref{tab:ism} shows the number of stars used per field. 
Figure \ref{fig:cfd} shows the results for six
sets of data: (a) using all the stars from the catalog, (b)--(f) using stars from region II to region V. For each set of data four CFDs were evaluated:
using the observed polarization and the foreground-corrected polarization considering the three aforementioned estimates.
The straight line at $45^{\circ}$ represents a random distribution. If a
distribution is above this line, $\theta_M$ is concentrated at smaller values, which indicates an alignment with the SMC--LMC direction.
A distribution below the straight line indicates the magnetic field being perpendicular to the SMC--LMC direction. 

	For all cases, barring region III, the $\theta_M$ distribution for the observed polarization is above the straight line, indicating an alignment with the SMC--LMC
direction. The foreground-corrected CFDs, considering all the different estimates, lie below the observed one. This correction brings the
distributions closer to the straight line. This demonstrates the importance of a proper foreground subtraction to understand a possible SMC magnetic field alignment with the SMC--LMC direction. This is
particularly true because the foreground polarization is approximately aligned with the SMC--LMC direction (Figure \ref{fig:maps}a). For all the
cases, except region IV--V, the distributions roughly follow the straight line until a noticeable deviation at around $40^{\circ}$. The location of the
deviation corresponds to the direction of trends I and III, as previously mentioned, at $65^{\circ}$ and $150^{\circ}$ ($\theta_{0M} + 40 - 90$
and $\theta_{0M} + 40$). As briefly touched
upon, region III does not follow the above explanation. In this case, the distributions, including the observed polarization,
lie slightly below the straight line. Similarly to the other regions a noticeable deviation up from the straight line can be seen at around $30^{\circ}$.
Region IV--V shows a similar behavior, but only for the foreground-corrected distribution.

	In order to check whether we made an appropriate choice for the definition of $\theta_{0M}$, we repeated the analysis this time adopting
$\theta_{0M}$ as the angle between the star's position and the LMC's center. Overall there is no qualitative difference in the results.

	To quantify by how much our CFDs are similar to the uniform one, we used the approach from \cite{bib:rodrigues09}. This involved doing a
Kuiper test, which is a variant of the Kolmogorov-Smirnov test and more
appropriate for cyclic quantities such as $\theta_M$. Most of the distributions had probabilities much smaller than $1\%$ of being uniform, except
for: (i) region IV using the \cite{bib:rodrigues97} foreground estimate, the probability is of $34\%$; (ii) region V using the \cite{bib:rodrigues97}
and this work foreground estimates, for which the probabilities are $8\%$ and $7\%$, respectively. The probabilities are not very high, therefore none of the cases can be
classified as a uniform distribution.

	Previous studies observed an overall alignment of the SMC's ordered magnetic field and the Magellanic Bridge direction.
Our sample shows such overall alignment only for the observed data. Once the foreground is removed a complex geometry for the magnetic field arises.
The CFDs analysis displays neither a uniform distribution nor a strong alignment with the Bridge direction. The SMC's magnetic field, despite
being complex, is not totally random. The new strong alignments correspond to trend I and III, while trend II is not predominant in the CFDs study.


\section{Magnetic Field Strength}\label{sec:Bintensity}

	Our data can also be used to estimate the magnetic field strength. For this purpose, we used the \cite{bib:cf53} method
modified by \cite{bib:falceta08}. Since the interstellar polarization is due to an alignment of the dust grains' angular momentum with a local magnetic field, one
can expect that for a larger magnetic field, the dispersion of the polarization angles will be smaller.

	Besides the polarization angles' dispersions ($\delta \theta$), which were already estimated using the Gaussian fits, we also need to know the gas
velocity dispersion in the line of sight direction ($\delta V_{LOS}$) and the gas mass density ($\rho$), in order to be able to apply this method.
These quantities were estimated using maps of HI velocity dispersion and HI column density from \cite{bib:stanimirovic99,bib:stanimirovic04}.

\subsection{HI Velocity Dispersion}\label{sec:HI}

	In our analysis we used a HI velocity dispersion map that was derived using combined images from the ATCA
and Parkes. The second-momentum analysis was used to obtain the velocity dispersion map, which can be seen in Figure 4 of \cite{bib:stanimirovic04}.

	Since this map includes just the SMC's body, $\delta V_{LOS}$ was locally estimated just for those fields coinciding with the map, {\it i.e.},
SMC01--19. The estimate for each field was obtained by averaging the line of sight velocities in the area of the $\delta V_{LOS}$ map
coinciding exactly with each of our 8~x~8~arcmin SMC fields. For fields located at the Magellanic Bridge, we considered the average
value for this region from \cite{bib:bruns05}: $\delta V_{LOS}~=~20$~km~s$^{-1}$.

\subsection{HI Mass Density}

	The same data set of Section \ref{sec:HI} was used. The HI velocity profiles were integrated to obtain the HI column density map, which is shown in Figure 3 of
\cite{bib:stanimirovic99}.

	We obtained the local estimates as previously described for fields SMC01--19. For the remaining fields, we used the average value of N(HI)~$=
5 \times 10^{20}$~atoms~cm$^{-2}$ \citep{bib:bruns05} for the Bridge. Before averaging over the squares to calculate N(HI), the
aforementioned quantity was corrected by a factor to account for the hydrogen auto-absorption along the line of sight and defined by the following
equation:
{\small
\begin{empheq}[left={f = \empheqlbrace}]{align}\centering
1 + 0.667(log~N(HI) - 21.4); & ~\text{for}~log N(HI) > 21.4, \nonumber\\
1; & ~\text{for}~log N(HI) \leq 21.4.
\label{eq:Habs}
\end{empheq}
}This correction is required for regions with column densities higher than $2.5 \times 10^{21}$~atoms~cm$^{-2}$ \citep{bib:stanimirovic99}.

	In order to convert from HI column density [$N(HI)$] to HI number density ($n_H$), the SMC's depth had to be estimated.
\cite{bib:subramanian12} used the dispersions in the magnitude-color diagrams of RC stars together with distance estimates of RRLS to determine the
SMC's depth along the line of sight. They obtained an average depth of $(14\pm6)$~kpc (error obtained by private communication with the first author). As mentioned in Section \ref{sec:intro}, the
line of sight depth at SMC varies from region to region, therefore adopting an average value leads to some uncertainty in the
determination of $n_H$. For the fields located at the Wing and Bridge, this average is a good estimate. For the fields at the NE, where the
depth is higher \citep{bib:nidever13}, we may be
underestimating this value, which leads to an overestimation of $n_H$. Nevertheless, the mass density was obtained by applying $\rho = \gamma m_H
n_H$. Considering the SMC's abundances the equivalent molecular weight is $\gamma = 1.22$ \citep{bib:mao08}.

\subsection{Magnetic Field Strength Estimates}

	Following \cite{bib:falceta08} method, we firstly determined the turbulent magnetic field strength ($\delta B$), assuming equipartition between the turbulent magnetic
field energy and turbulent gas kinetic energy (Equation \ref{eq:dB}). Lastly, we determined the ordered magnetic field strength on the plane of
the sky ($B_{sky}$) through Equation \ref{eq:Bsky}.

\begin{empheq}{align}\centering
& \frac{1}{2}\rho\delta V_{\text{LOS}}^2 \simeq \frac{1}{8\pi}\delta B^2, \label{eq:dB}\\
& B_{\text{sky}} + \delta B \simeq \sqrt{4\pi\rho}\frac{\delta V_{\text{LOS}}}{tan(\delta\theta)}. \label{eq:Bsky}
\end{empheq}

	Table \ref{tab:ism} shows the obtained magnetic field intensities for each field, as well as the values for the HI velocity dispersion at the
line of sight and HI number
density. For the fields with more than one PD, different estimates were calculated using the different polarization angles' dispersion,
which are also shown in Table \ref{tab:ism}. When we obtain more than one PD for the polarization angle dispersion, we know that they must be
located at different distances or direction and for the first case the values of $\delta V_{LOS}$ and $n_H$ would be more appropriate for the most distant
PD. It is not easy to quantify the uncertainty for our estimates, but we want to point out that the errors might be quite high. Nevertheless, the
usage of the integrated values lead to an overestimation of $n_H$ and $\delta V_{LOS}$, therefore overestimating $\delta B$ and underestimating
$B_{\text{sky}}$.

\begin{table*}[!htb]\centering
\caption{Local ISM parameters for our fields}
\begin{threeparttable}
\begin{tabular}{cccccccccc}\toprule\toprule

SMC\tnote{a} & $\delta V_{LOS}$\tnote{b} & $n_H$\tnote{c} & $\delta \theta$\tnote{d} &
$\delta B$\tnote{e} & $B_{{sky}}$\tnote{f} & $\theta$\tnote{g} & $P$\tnote{h} &
Label\tnote{i} & No.\tnote{j}\\

 & $(\text{km~s}^{-1})$ & $(\text{atoms~cm}^{-3})$ & $(\text{deg})$ &
$(\mu \text{G})$ & $(\mu \text{G})$ & $(\text{deg})$ & $(\%)$ &
  & \\\midrule

\multirow{2}{*}{01} & \multirow{2}{*}{$12.643\pm0.078$} & \multirow{2}{*}{$0.102\pm0.043$}   & $21.28 \pm 0.29$ &
\multirow{2}{*}{$2.05\pm0.44$} & $3.21\pm0.69$  & $      120.90  \pm     0.28    $      &       $0.90$    &
\multirow{2}{*}{P2} & \multirow{2}{*}{491}\\
 & & & $22.9 \pm  1.2$ & & $2.80\pm0.66$ & $      46.3    \pm     1.1$ & $0.85$ & \\ 

\multirow{2}{*}{02} & \multirow{2}{*}{$14.763\pm0.091$} & \multirow{2}{*}{$0.092\pm0.039$}   & $40.93 \pm 0.90$ & 
\multirow{2}{*}{$2.27\pm0.48$} & $0.35\pm0.11$  & $	58.39	\pm	0.73	$	&	$0.62$	&
\multirow{2}{*}{P2} & \multirow{2}{*}{492}\\
 & & & $19.5 \pm 2.2$ & & $4.1\pm1.2$ & $      150.0   \pm     1.6$ & $0.66$ & \\

03 & $24.228\pm0.075$ & $0.144\pm0.061$     & $46.08 \pm 0.61$ & $4.66\pm0.99$ 	& $-0.17\pm0.10$ &	  
$	102.16	\pm	0.55	$	&	$1.0$	& P1 & 471 \\         

\multirow{2}{*}{04} & \multirow{2}{*}{$25.723\pm0.072$} & \multirow{2}{*}{$0.183\pm0.078$}     & $39.7 \pm 1.9$ & \multirow{2}{*}{$5.6\pm1.2$}
& $1.14\pm0.51$  &	  
$	169.2	\pm	1.6	$	&	$0.87$	& \multirow{2}{*}{P2} & \multirow{2}{*}{170}\\             
 & & & $10.8 \pm 2.2$ & & $23.6\pm7.9$ & $      65.5    \pm     2.0$   &       $1.0$ & \\

05 & $13.799\pm0.093$ & $0.174\pm0.074$     & 	-- 	   & $2.92\pm0.62$ & -- 		   &	  
		--			&			--		& P0 & 47 \\             

06 & $19.690\pm0.085$ & $0.164\pm0.070$     & $71.9 \pm 3.2$   & $4.04\pm0.86$ & $-2.72\pm0.63$ &	  
$	100.6	\pm	2.1	$	&	$1.2$	& P1 & 247 \\              

\multirow{2}{*}{07} & \multirow{2}{*}{$27.824\pm0.045$} & \multirow{2}{*}{$0.165\pm0.070$}     & $43.0 \pm 2.2$ & \multirow{2}{*}{$5.7\pm1.2$}
& $0.42\pm0.48$  &	  
$	84.7	\pm	1.6	$	&	$1.0$	& \multirow{2}{*}{P2} & \multirow{2}{*}{640}\\             
 & & & $25.28 \pm 0.72$ & & $6.4\pm1.4$ & $      178.85  \pm     0.73$  &       $1.0$ & \\

\multirow{2}{*}{08} & \multirow{2}{*}{$22.09\pm0.13$} 	& \multirow{2}{*}{$0.104\pm0.044$}   & $37.03 \pm 0.51$ &
\multirow{2}{*}{$3.61\pm0.77$} & $1.18\pm0.26$  &	  
$	73.69	\pm	0.38	$	&	$0.87$	& \multirow{2}{*}{P2} & \multirow{2}{*}{735}\\             
 & & & $21.76 \pm 0.88$ & & $5.4\pm1.2$ &  $      154.79  \pm     0.66$  &       $0.93$ & \\

09 & $26.332\pm0.066$ & $0.154\pm0.066$     & $18.84 \pm 0.16$ & $5.2\pm1.1$ 	& $10.1\pm2.2$   &	  
$	67.51	\pm	0.16	$	&	$0.98$	& P1 & 605 \\             

\multirow{2}{*}{10} & \multirow{2}{*}{$23.764\pm0.063$} & \multirow{2}{*}{$0.154\pm0.066$}     & $26.45 \pm 0.36$ &
\multirow{2}{*}{$4.7\pm1.0$} 	& $4.8\pm1.0$    &	  
$	142.54	\pm	0.31	$	&	$0.94$	& \multirow{2}{*}{P2} & \multirow{2}{*}{560}\\             
 & & & $22.67 \pm 0.72$ & &  $6.6\pm1.5$ & $      54.97   \pm     0.60$  &       $1.0$ & \\

11 & $22.443\pm0.071$ & $0.160\pm0.068$     & $17.17 \pm 0.17$ & $4.55\pm0.97$ 	& $10.2\pm2.2$   &  
$	158.29	\pm	0.17	$	&	$1.4$	& P1 & 453 \\             

12 & $22.200\pm0.049$ & $0.083\pm0.035$   & $14.55 \pm 0.68$   & $3.24\pm0.69$ 	& $9.3\pm2.1$   &  
$	147.88	\pm	0.68	$	&	$0.74$	& PFM & 205 \\           

13 & $19.89\pm0.14$ 	& $0.163\pm0.069$     & $24.1 \pm 2.6$   & $4.07\pm0.86$ 	& $5.0\pm1.5$    &	  
$	126.2	\pm	2.5	$	&	$0.46$	& PFM & 122 \\          

14 & $21.08\pm0.14$ 	& $0.137\pm0.058$     & 	-- 	   & $3.95\pm0.84$ 	& -- 		   &	  
		--			&	--	& P0 & 165 \\          

15 & $23.31\pm0.11$ 	& $0.134\pm0.057$     & $27.2 \pm 1.1$   & $4.32\pm0.92$ 	& $4.08\pm0.96$    &	  
$	144.5	\pm	1.0	$	&	$1.1$	& P1 & 139 \\         

\multirow{2}{*}{16} & \multirow{2}{*}{$20.77\pm0.12$} 	& \multirow{2}{*}{$0.056\pm0.024$}   & $25.6 \pm 4.6$ &
\multirow{2}{*}{$2.48\pm0.53$} & $2.7\pm1.2$    &	  
$	134.5	\pm	4.8	$	&	$0.89$	& \multirow{2}{*}{PSE} & \multirow{2}{*}{87}\\            
 & & & $22.8 \pm 7.0$ & & $3.4\pm2.1$ &  $      66.8    \pm     6.4$   &       $0.72$ & \\

\multirow{2}{*}{17} & \multirow{2}{*}{$20.602\pm0.069$} & \multirow{2}{*}{$0.097\pm0.041$}   & $30.0 \pm 3.2$ & \multirow{2}{*}{$3.26\pm0.69$}
& $2.38\pm0.89$    &	  
$	118.9	\pm	2.6	$	&	$0.87$	& \multirow{2}{*}{PSE} & \multirow{2}{*}{136}\\            
 & & & $15.0 \pm 1.6$ & & $8.9\pm2.3$ & $      54.0    \pm     1.6$   &       $0.92$ & \\

18 & $20.842\pm0.084$ & $0.059\pm0.025$   & $59.5 \pm 3.7$   & $2.57\pm0.55$ & $-1.06\pm0.32$ &	  
$	88.5	\pm	2.9	$	&	$1.1$	& P1 & 131 \\             

19 & $18.96\pm0.14$ 	& $0.046\pm0.019$   & 	-- 	   & $2.05\pm0.44$ & -- 		   &	  
		--			&	--	& P0 & 40 \\            

20 & $20$ 		& $0.0115\pm0.0049$ & $20.6 \pm 3.2$   & $1.09\pm0.23$ & $1.80\pm0.62$  &	  
$	75.0	\pm	3.2	$	&	$0.69$	& PFM & 63 \\          

21 & $20$ 		& $0.0115\pm0.0049$ & $7.51 \pm 0.42$  & $1.09\pm0.23$ & $7.1\pm1.6$   &	  
$	159.57	\pm	0.49	$	&	$1.7$	& P1 & 57 \\          

\multirow{2}{*}{22} & \multirow{2}{*}{$ 20$} 		& \multirow{2}{*}{$0.0115\pm0.0049$} & $35.2 \pm 5.4$ &
\multirow{2}{*}{$1.09\pm0.23$} & $0.45\pm0.32$  &	  
$	157.8	\pm	4.3	$	&	$0.96$ & \multirow{2}{*}{P2} & \multirow{2}{*}{60}\\       
 & & & $21 \pm 15$ & & $1.7\pm2.2$ &  $      75      \pm     14$    &       $0.89$ & \\

23 & $20$ 		& $0.0115\pm0.0049$ & $8.9 \pm 1.3$   & $1.09\pm0.23$ & $5.8\pm1.6$  &	  
$	158.8	\pm	1.2	$	&	$0.75$	& PFM & 63 \\            

24 & $20$ 		& $0.0115\pm0.0049$ & $27.9 \pm 2.4$   & $1.09\pm0.23$ & $0.96\pm0.29$  &	  
$	150.1	\pm	2.4	$	&	$1.3$	& PSE & 63 \\           

25 & $20$ 		& $0.0115\pm0.0049$ & 	-- 	   & $1.09\pm0.23$ & -- 		   &	  
		--			&	--	& P0 & 40 \\           

26 & $20$ 		& $0.0115\pm0.0049$ & $31.3 \pm 3.2$   & $1.09\pm0.23$ & $0.70\pm0.27$  &	  
$	79.4	\pm	2.9	$	&	$0.92$	& PSE & 60 \\         

\multirow{2}{*}{27} & \multirow{2}{*}{$20$} 		& \multirow{2}{*}{$0.0115\pm0.0049$} & $22.7 \pm 2.9$ &
\multirow{2}{*}{$1.09\pm0.23$} & $1.51\pm0.49$  &	  
$	69.4	\pm	2.6	$	&	$0.69$	& \multirow{2}{*}{P2} & \multirow{2}{*}{72}\\             
 & & & $7.8 \pm 1.2$ & & $6.8\pm1.9$ & $      116.2   \pm     1.0$   &       $0.67$   & \\

28 & $20$ 		& $0.0115\pm0.0049$ & $13.90 \pm 0.77$ & $1.09\pm0.23$ & $3.30\pm0.74$    &	  
$	69.87	\pm	0.77	$	&	$0.82$	& P1 & 72 \\\bottomrule  

\end{tabular}

\begin{tablenotes}
{\footnotesize
\item[a] Field's label. 
\item[b] HI velocity dispersion.
\item[c] HI number density.
\item[d] Polarization angle dispersion, for fields with two PDs the first row shows the first PD and the second row the second PD.
\item[e] Turbulent magnetic field.
\item[f] Ordered sky-projected magnetic field, for fields with two PDs the first row shows the first PD and the second row the second PD.
\item[g] Trend for the polarization angle, for fields with two PDs the first row shows the first PD and the second row the second PD. 
\item[h] Median polarization, for fields with two PDs the first row shows the first PD and the second row the second PD.
\item[i] Field's classification.
\item[j] Number of stars with $P/\sigma_P > 3$.
}

\end{tablenotes}
\end{threeparttable}
\label{tab:ism}
\end{table*}

	The fields SMC 03, 06, and 18 have negative estimates for $B_{sky}$. The fact that we may be underestimating $B_{sky}$ can be an explanation
for the negative values; nonetheless, these fields are also the ones with larger $\delta \theta$, all of which
are higher than
$\pi/4$. For dispersions of that order, the magnetic field may have just a turbulent component or a high inclination with respect to the
plane of the sky \citep{bib:falceta08}. We know from Faraday rotation that there is an ordered line of sight magnetic field of
$B_{LOS}~=~(0.19~\pm~0.06)~\mu$~G \citep{bib:mao08}, despite this small value there is the possibility that in these regions $B_{LOS}$ dominates over
$B_{sky}$. It should also be noted that all the values of 
$B_{sky}$ (with exception of the negative ones), and certainly the average $B_{\text{sky}}$ value, are significantly higher than the average
$B_{LOS}$. This shows that the magnetic field of the SMC is, in general, mostly in the plane of the sky. This lends additional value to the
study of the $B_{\text{sky}}$ structure in the SMC.

\begin{figure}[!htp]\centering
\figurenum{10}
\includegraphics[width=1.\columnwidth]{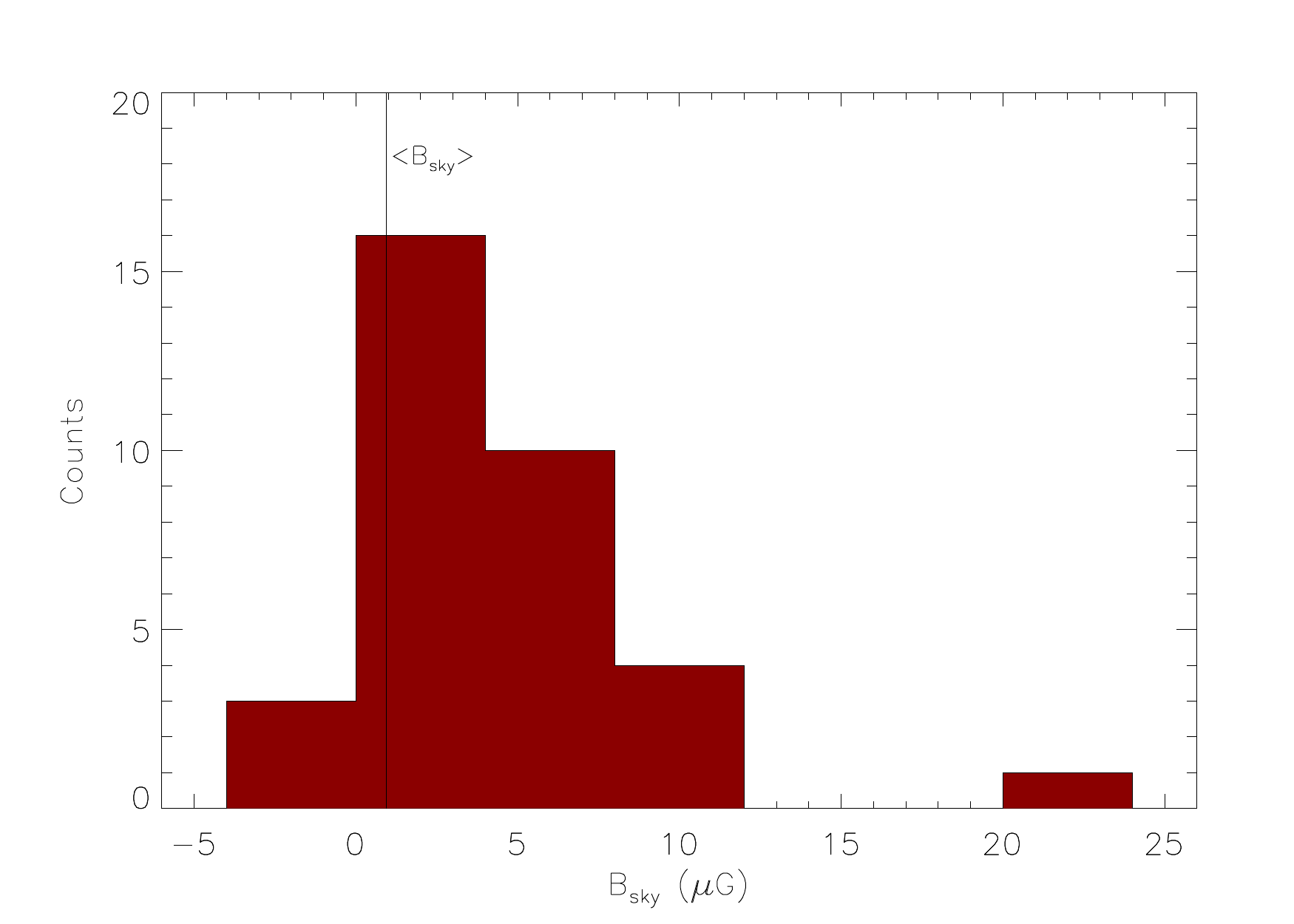}
\caption{Histogram of the ordered sky-projected magnetic field. The vertical line at $0.947~\mu$G represents the uncertainty weighted average
(excluding the negative values).}
\label{fig:Bsky}
\end{figure}

	The $B_{sky}$ dispersion is quite high as can be seen in Figure \ref{fig:Bsky}. In many cases, the values can be up to $25$ times above the
average. This high dispersion is not so surprising, because our estimates may contain not just
measurements of the diffuse interstellar magnetic field, but also estimates for local structures (e.g., shells, clusters, molecular clouds,
HII regions). Optical polarimetric observations already showed that some ISM structures are magnetized, possessing magnetic fields of up
to tens of $\mu$G, for instance, IRAS Vela Shell \citep{bib:pereyra07} and NGC 2100 \citep{bib:wisniewski07}, so that some of the high values of
$B_{sky}$ could be associated to such structures. To verify this hypothesis we queried Simbad\footnote[8]{\url{http://simbad.u-strasbg.fr/simbad/}}
in the regions of our fields and looked for ISM structures. Table \ref{tab:ismobjs} presents
a list of objects per field. Some of the
measurements may be related to these structures; nonetheless, we can only guarantee that there is spatial projected correlation between the structures and the
polarization vectors. With respect to the fields that possess magnetic fields of tens of $\mu$G, SMC04, for instance, possesses a shell with a size of
$5$x$5$~arcmin, this structure might be responsible for our estimated value for $B_{\text{sky}}=23.6~\mu$G. The fields with a large
$\delta\theta$, SMC 06 and 14, for instance, are the fields with a largest number of structures. It is possible to find direct associations of the polarization angles with the geometry of these structures, but for doing this a detailed study
of each source would be necessary, which is beyond the scope of this work.

\begin{table*}[!htb]\centering
\caption{ISM structures in our fields}
\begin{threeparttable}
\begin{tabular}{cllll}\toprule\toprule

SMC\tnote{a} & Object Name & R.A.\tnote{b} & Decl.\tnote{b} & Object Type \\
 & & (h:m:s) & ($\arcdeg$:$\arcmin$:$\arcsec$) &  \\\midrule

01	&	DEM S 107 		&	01:00:08.7	&	-71:48:04	&	HII region	\\
	&	[MOH2010] NE-1c 	&	01:00:10.0	&	-71:48:40	&	molecular cloud	\\
	&	DEM S 113		&	01:01:30.3	&	-71:47:45	&	HII region \\
	&	[SSH97] 290 		&	01:01:12  	&	-71:52.9 	&	shell 	\\
	&	LHA 115-N 72 		&	01:01:32.5	&	-71:50:43	&	HII region	\\
\hline
02	&	[SSH97] 279	 	&	01:00:26 	&	-72:05.1 	&	shell 	\\
	&	[SSH97] 285 		&	01:00:46 	&	-72:09.6 	&	shell 	\\
\hline
03	&	[SSH97] 288 		&	01:01:09 	&	-72:24.6 	&	shell 	\\
\hline
04	&	[JD2002] 21 		&	01:01:33.6 	&	-72:34:53	&	planetary nebula 	\\
	&	[SSH97] 273 		&	01:00:05 	&	-72:39.7 	&	shell 	\\
\hline
05	&	DEM S 121 		&	01:03:03 	&	-71:53.8 	&	HII region	\\
	&	[SSH97] 319 		&	01:04:40 	&	-71:53.8 	&	shell 	\\
\hline
06	&	[MOH2010] NE-3g 	&	01:03:10.0 	&	-72:03:50 	&	molecular cloud	\\
	&	DEM S 119 		&	01:03:01.9 	&	-72:05:17 	&	HII region	\\
	&	LHA 115-N 76B 		&	01:03:08.0 	&	-72:06:24 	&	HII region	\\
	&	LHA 115-N 76A 		&	01:03:48.9 	&	-72:03:52 	&	HII region	\\
	&	[SSH97] 309 		&	01:03:03   	&	-72:03.2  	&	shell 	\\
	&	SNR B0101-72.6 		&	01:03:17   	&	-72:09.7  	&	supernova remnant 	\\
	&	MCELS S-148		&	01:03:48.6 	&	-72:03:56	&	HII region \\
	&	MCELS S-147		&	01:03:25.0 	&	-72:03:45	&	HII region \\
	&	MCELS S-144		&	01:03:01.2 	&	-72:05:41	&	HII region \\
\hline
07	&	[SSH97] 329		&	01:05:18	&	-72:34.4	&	shell	\\
	&	[SSH97] 330		&	01:05:23	&	-72:38.9	&	shell	\\
	&	[SSH97] 321		&	01:04:42	&	-72:37.5	&	shell	\\
\hline
08	&	DEM S 134 		&	01:06:48.5 	&	-72:24:32	&	HII region	\\
	&	[SSH97] 344 		&	01:06:46 	&	-72:25.0 	&	shell 	\\
	&	[SSH97] 355 		&	01:07:30 	&	-72:21.8 	&	shell 	\\
\hline
09	&	[SSH97] 367 		&	01:08:43  	&	-72:38.1 	&	shell 	\\
	&	[SSH97] 358 		&	01:07:33  	&	-72:39.9 	&	shell 	\\
\hline
10	&	[SSH97] 354		&	01:07:25	&	-72:54.9	&	shell	\\
	&	DEM S 133		&	01:07:34.7	& 	-72:51:20	&	HII region \\
\hline
11	&	[SSH97] 376		&	01:09:29	&	-72:57.4	&	shell	\\
	&	[SSH97] 386		&	01:10:31	&	-73:03.1	&	shell	\\
	&	[SSH97] 373		&	01:09:20	&	-73:01.9	&	shell	\\
	&	[SSH97] 387		&	01:10:33	&	-73:01.6	&	shell	\\
\hline
12	&	[BLR2008] SMC N83 4	&	01:12:05.8	&	-73:31:01	&	molecular cloud	\\
	&	[SSH97] 398 	   	&	01:12:23  	&	-73:28.0 	&	shell 	\\
	&	[BLR2008] SMC N83 2	&	01:12:41.3	&	-73:32:14	&	molecular cloud	\\
\hline
13	&	[SSH97] 408		&	01:13:30	&	-73:03.6	&	shell	\\
	&	[SSH97] 407		&	01:13:29	&	-73:03.6	&	shell	\\
\hline
14	&	2MASX J01144713-7320137 &	01:14:47.132 	&	-73:20:13.80	&	planetary nebula 	\\
	&	DEM S 152 		&	01:14:54.1 	&	-73:19:45   	&	HII region	\\
	&	NAME SMC B0113-7334 	&	01:14:44.9 	&	-73:20:06   	&	HII region	\\
	&	DEM S 157 		&	01:16:20 	&	-73:20.2    	&	HII region	\\
	&	[SSH97] 423 		&	01:15:29 	&	-73:23.8    	&	shell 	\\
	&	MCELS S-193		&	01:14:55.7 	&	-73:20:10	&	HII region \\
	&	MCELS S-195		&	01:15:04.7 	&	-73:19:10	&	HII region \\
	&	MCELS S-191		&	01:14:41.7 	&	-73:18:06	&	HII region \\
\hline
15	&	[SSH97] 440 		&	01:17:27	&	-73:12.5	&	shell 	\\
	&	DEM S 159 		&	01:16:58	&	-73:12.1	&	HII region	\\
	&	DEM S 155 		&	01:17.1 	&	-73:14 	&	HII region	\\
\hline
17	&	LHA 115-N 87		&	01:21:10.69 	&	-73:14:34.8	&	planetary nebula	\\
	&	[SSH97] 472 		&	01:21:40 	&	-73:18.0   	&	shell	\\
\hline
19	&	[MSZ2003] 28		&	01:30:44 	&	-73:49:42	&	shell \\
	&	[MSZ2003] 32		&	01:31:43 	&	-73:52:24	&	shell \\
\hline
20	&	[MSZ2003] 49		&	01:42:36 	&	-73:49:54	&	shell \\
	&	[MSZ2003] 48		&	01:41:35 	&	-73:55:12	&	shell \\
\hline
21	&	[MSZ2003] 52		&	01:45:23 	&	-74:30:06	&	shell \\
	&	[MSZ2003] 55		&	01:46:10 	&	-74:28:24	&	shell \\
\hline
26	&	[MSZ2003] 88		&	02:06:37 	&	-74:36:48	&	shell \\
\hline
27	&	[MSZ2003] 95		&	02:09:22 	&	-74:24:48	&	shell \\
\bottomrule

\end{tabular}

\begin{tablenotes}
{\footnotesize
\item[a] Field's label.
\item[b] Coordinates in J2000.
}

\end{tablenotes}
\end{threeparttable}
\label{tab:ismobjs}
\end{table*}

	A turbulent magnetic field value of $\delta B=~(1.465~\pm~0.069)~\mu$G was obtained from the uncertainty weighted average, when considering
all fields. The same analysis, including both observed components, led to the computation of the ordered magnetic field projected on the plane of the sky:
$B_{sky}~=~(0.947~\pm~0.079)~\mu$G. The negative values were not used to evaluate the average. Our estimate for $\delta B$ is about $30\%$ and $60\%$
smaller than the value obtained by \cite{bib:mao08} and \cite{bib:magalhaes09}, respectively. The trend of smaller values is also observed in the
results for $B_{sky}$, which is $40\%$ smaller than \cite{bib:mao08} and $50\%$ reduced relative to \cite{bib:magalhaes09}. The high difference in these estimates arise mainly due to the way $n_H$ was evaluated. Both
\cite{bib:mao08} and \cite{bib:magalhaes09} used a constant value for the HI number density ($n_H = 0.1$~atoms~cm$^{-3}$). We can see in Table
\ref{tab:ism} that the fields SMC20--28 have $n_H$ one order of magnitude smaller than the value quoted in the previously mentioned papers, therefore
our estimates for the magnetic field in these regions should naturally lead to smaller values, also reducing the average. The observation of a
weak large scale magnetic field agrees with the expectation that the HI in the SMC is super-Alfv\'enic \citep{bib:burkhart10}.

	The average turbulent component is higher than
the average ordered component, a common result for all types of galaxies. The $B_{sky}/\delta B = 0.65$ ratio for the SMC is closer to the MW, ranging
from $0.6$ to $1.0$ \citep{bib:beck01}, than the typical values for other irregular dwarfs, $\sim 0.2$ \citep{bib:chyzy11}.
The production of magnetic fields in irregular dwarf galaxies is most likely not maintained by
a large-scale dynamo process, due to the small rates observed for $B_{sky}/\delta B$ \citep{bib:chyzy11}. In the case of the SMC, this process can be
the mechanism that creates the large-scale magnetic field observed in the SMC, as discussed in Section \ref{sec:Bgeometry}.


\section{Summary and Conclusions}\label{sec:conclusion}

	This work used optical polarimetric data from CTIO, aiming to study the magnetic field of the SMC, an irregular galaxy and satellite of the
MW. One of the biggest peculiarities of the SMC is that its ISM is particularly different from
that of the
Galaxy (e.g., high gas-to-dust ratio and submm excess emission), most likely due to the large difference in metalicity. The data reduction led to a catalog with 7207 stars, with well
determined polarizations ($P/\sigma_P\geq3$). This new catalog is a great improvement compared to previous catalogs for the SMC.
 
	Our analysis showed that caution is necessary when subtracting the foreground Galactic polarization, because this correction strongly changes the geometry observed for the magnetic
field. We present a new estimate for the foreground Galactic polarization using the stars from our catalog, which has smaller errors compared to
previous ones.

	This catalog was used to study the magnetic field on the SMC. After foreground removal, three trends at the following polarization angles were
observed: ($65^{\circ}\pm10^{\circ}$), ($115^{\circ}\pm10^{\circ}$), and ($150^{\circ}\pm10^{\circ}$). For the first trend, the polarization vectors
in the NE region are roughly aligned with the Bar direction, which is at a PA of $\sim45^{\circ}$,
reinforcing what was observed by \cite{bib:magalhaes09}. In the case of the second trend, the polarization angle is aligned with the
Bridge direction, which is at a PA of $115^{\circ}.4$, and possess the same direction as the Galactic foreground, which may question its veracity. This trend has been seen and confirmed in many studies
\citep{bib:schmidt70,bib:mf70a,bib:mf70b,bib:schmidt76,bib:magalhaes90,bib:mao08,bib:magalhaes09}. A possible explanation for the magnetic field
alignment with the Bar direction is that the magnetic field is coupled to the gas, therefore the field lines are parallel to the Bar direction due to
the flux freezing condition. The
second trend may be due to tidal stretching of the magnetic field lines in the direction of the Magellanic Bridge. The coincidence of the alignment of the second trend
with the Galactic foreground may be due to the Galactic halo also feeling the tides from the MCs in the region close to the MCs, therefore
the MW's halo magnetic field gets also stretched in the same direction. The third trend does not display any particular feature. Regardless the trends, the
magnetic field structure seems to be rather complex in the SMC.

	The polarization and magnitude distributions of the $115^{\circ}$ and $150^{\circ}$ trends suggest that the former is located
closer to us, with the latter located further away. Distances of Cepheids show a bimodality \citep{bib:mfv86,bib:nidever13} that is also
observed in the HI
velocities \citep{bib:mathewson84,bib:stanimirovic04}. Hence, each of our trends may be related to a different component. The trend at $65^{\circ}$
is most likely present in the two components of the SMC.

	We obtained a turbulent component for the magnetic field of $\delta B~=~(1.465\pm0.069)~\mu$G and for the ordered magnetic field projected on the plane
of the sky of $B_{sky}~=~(0.947\pm0.079)~\mu$G. The ordered-to-random field ratio at the SMC is closer to what is observed in our Galaxy than
the average values for other irregular dwarf galaxies.

	This study is relevant for a better understanding of the magnetic field at the SMC, with a catalog containing good polarization
determinations,
which can be used for several kinds of studies. We wish to emphasize that our data were concentrated at the NE and Wing sections of the SMC and part
of the Magellanic Bridge. Further observations including the whole SMC, LMC, and Magellanic Bridge are necessary for a more complete picture of this
system. The forthcoming SOUTH POL project data \citep{bib:southpol} will map the polarization for the entire southern sky. It will cover
the entire SMC--LMC system and Magellanic Stream and Bridge, which will allow a better understanding of the spatial magnetic field
behavior. SOUTH POL is expected to measure the polarizations of around 24,500 bright stars ($m_V \lesssim 15$~mag) in this system
with accuracy up to $0.1\%$ and around 218,000 stars ($m_V \lesssim 17$~mag) with accuracy up to $0.3\%$.


~
~

\acknowledgements

ALG is grateful for the support of the International Max Planck Research School at Heidelberg (IMPRS-HD), {\it Funda\c{c}\~ao de Amparo \`a Pesquisa
do Estado de S\~ao Paulo} (FAPESP, process no. 2010/03802-2), and {\it
Conselho Nacional de Desenvolvimento Cient\'ifico e Tecnol\'ogico} (CNPq, process no. 132834/2010-3). AMM is
grateful for support by FAPESP (grants nos.
01/12589-1, 2001/12589-1, and 2010/19694-4) and CNPq's Research Grant. CVR is grateful for support by CNPq, grant no. 306103/2012-5. We also would like to thank Snezana Stanimirovi\'c, who
kindly provided us with the HI maps; Smitha Subramaniam for the insights regarding the SMC's depth; Benjamin Laevens for careful reading of the
text, which highly improved its quality; and Reinaldo Santos-Lima for the comments regarding the magnetic field geometry interpretation. Additionally,
we are very grateful for the helpful and fruitful comments by the anonymous referee.


\appendix

\section{The Polarimetric catalogs}\label{sec:appendixa}

	Here we present a short version of the polarimetric catalogs.

\begin{table*}[!htb]\centering
\caption{Polarimetric catalog (observed polarization)}
\begin{threeparttable}
\begin{tabular}{cccccccc}\toprule\toprule

ID & R.A. & Decl. & $P_{obs}$ & $\sigma_{P_{obs}}$ & PA & $V$ & $\sigma_V$ \\ 
 & (h:m:s) & ($\arcdeg$:$\arcmin$:$\arcsec$) & ($\%$) & ($\%$) & (deg) & (mag) & (mag) \\\midrule

0001 & 1:00:05.29 & -71:49:46.23 & 0.4120 & 0.1330 & 163.77 & 18.58280 & 0.16040 \\
0002 & 1:00:04.48 & -71:52:41.15 & 0.9270 & 0.0490 & 110.77 & 16.71380 & 0.16000 \\
0003 & 1:00:03.63 & -71:54:54.25 & 0.6460 & 0.1050 & 057.57 & 17.59820 & 0.16000 \\
0004 & 1:00:05.28 & -71:51:45.71 & 0.8860 & 0.1710 & 050.37 & 17.31070 & 0.16000 \\
0005 & 1:00:05.76 & -71:50:21.05 & 0.8300 & 0.0930 & 158.07 & 18.65590 & 0.16020 \\\bottomrule

\end{tabular}

\begin{tablenotes}
{\footnotesize
\item Table 7 is published in its entirety in the electronic edition of the {\it Astrophysical Journal}.  A portion is shown here for guidance regarding its form and content.
}
\end{tablenotes}

\label{tab:Pobs}
\end{threeparttable}
\end{table*}

\begin{table*}[!htb]\centering
\caption{Polarimetric catalog (intrinsic polarization)}
\begin{threeparttable}
\begin{tabular}{cccccccc}\toprule\toprule

ID & R.A. & Decl. & $P_{int}$ & $\sigma_{P_{int}}$ & PA & $V$ & $\sigma_V$ \\ 
 & (h:m:s) & ($\arcdeg$:$\arcmin$:$\arcsec$) & ($\%$) & ($\%$) & (deg) & (mag) & (mag) \\\midrule

0001 & 1:00:05.29 & -71:49:46.23 & 0.5781 & 0.1339 & 179.78 & 18.58280 & 0.16040 \\
0002 & 1:00:04.48 & -71:52:41.15 & 0.6109 & 0.0515 & 110.33 & 16.71380 & 0.16000 \\
0003 & 1:00:03.63 & -71:54:54.25 & 0.8027 & 0.1062 & 046.57 & 17.59820 & 0.16000 \\
0004 & 1:00:05.28 & -71:51:45.71 & 1.0892 & 0.1717 & 043.28 & 17.31070 & 0.16000 \\
0005 & 1:00:05.76 & -71:50:21.05 & 0.9030 & 0.0943 & 168.31 & 18.65590 & 0.16020 \\\bottomrule

\end{tabular}

\begin{tablenotes}
{\footnotesize
\item Table 8 is published in its entirety in the electronic edition of the {\it Astrophysical Journal}.  A portion is shown here for guidance regarding its form and content.
}
\end{tablenotes}

\label{tab:Pint}
\end{threeparttable}
\end{table*}

\section{Classification of the fields}\label{sec:appendixb}

	We used the $F$-test to quantify whether the number of Gaussians used to fit each of our fields was appropriate to represent the data set.
The $F$-test applied for regression problems was used. We assessed whether a model with more parameters (one more Gaussian in our case) would fit
the data better. The $F$ statistic is given by
\begin{equation}
F = \frac{\Big ( \frac{RSS_1 - RSS_2}{p_2 - p_1} \Big )}{\Big ( \frac{RSS_2}{n - p_2} \Big )},
\end{equation}
where $RSS_i$ is the residual sum of squares of model $i$, $p_i$ is the number of parameters of model $i$, and $n$ is the number of points used to fit
the data.

	$F$ has an $F$ distribution with ($p_2-p_1$, $n-p_2$) degrees of freedom. The null hypothesis of our test is that model 2 does not fit the data
better than model 1. We assumed a significance level probability of $2.5\%$, which gives statistically significant results, and looked for the critical values of $F$ in an online $F$ distribution
table\footnote[9]{\url{http://www.socr.ucla.edu/applets.dir/f_table.html}}. The null hypothesis is rejected if the $F$ obtained from the data is greater
than the critical value. Table \ref{tab:ftest} summarizes our results.

	The $F$-test demonstrates that all fields were well classified. The negative values of $F$ indicate that the RSS of model 2 is greater than that of
model 1, which by itself already shows that model 1 fits the data better. The fields SMC 08, 10, 14, 25, 26, and 27 have limiting values of $F$. In the
case where $RSS_2 >> RSS_1$ such that $RSS_1/RSS_2 << 1$, $F$ simplifies to

\begin{empheq}[]{align}\centering
 & F = \frac{(n-p_2)}{(p_2-p_1)}\Big ( \frac{RSS_1}{RSS_2} - 1\Big ) \nonumber \\
 & F \simeq -\frac{(n-p_2)}{(p_2-p_1)}. \nonumber
\end{empheq}

	The six aforementioned fields satisfied this limit, indicating that the simpler model is the most robust.

\section{Definition of the limiting magnitude for the random background}\label{sec:appendixc}

\begin{figure*}[!htb]\centering
\figurenum{11}
\includegraphics[width=1.\textwidth]{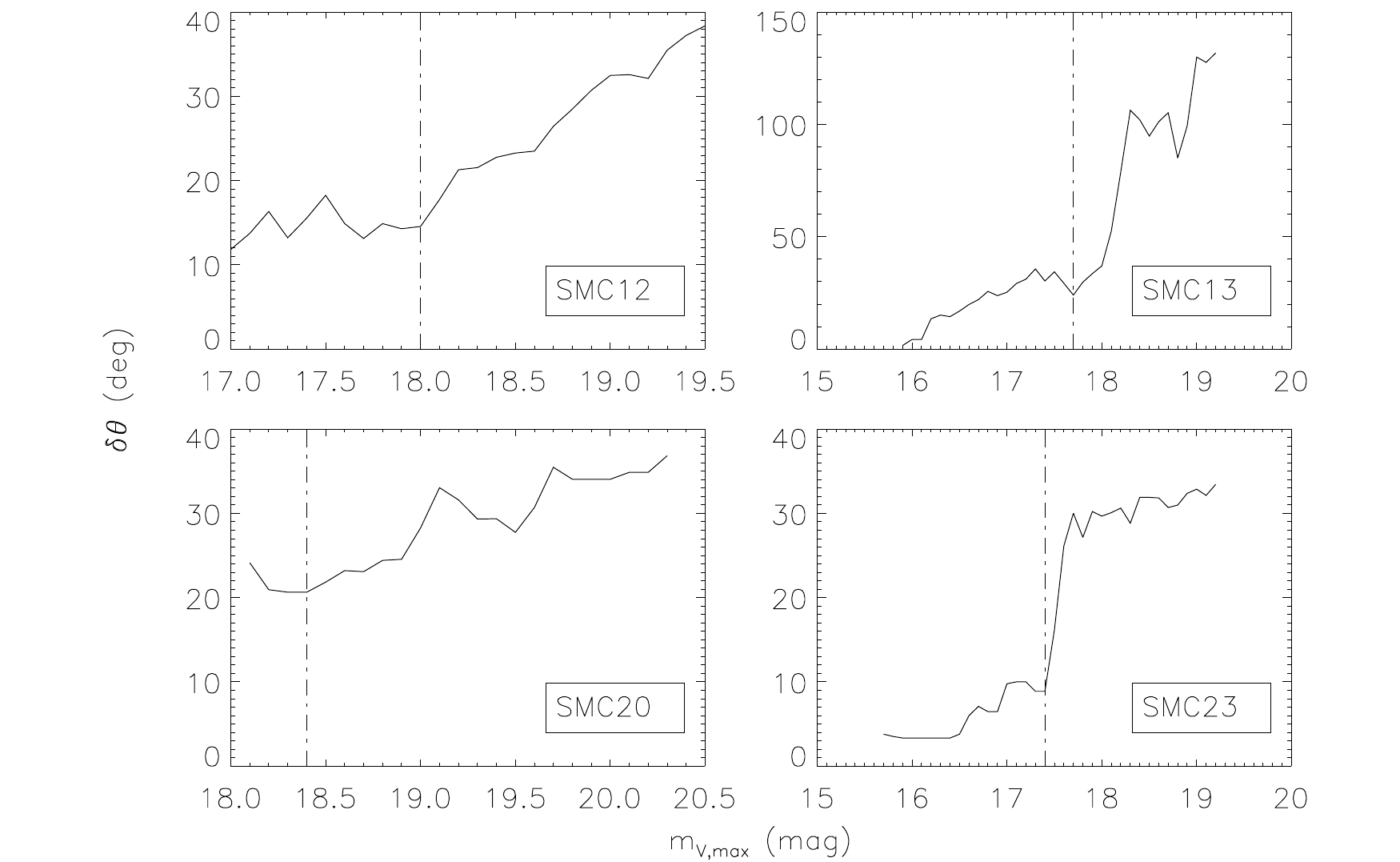}
\caption{Plots of the dispersion of $\theta$ as a function of the maximum magnitude considered. The dot-dashed straight line defines the limiting
magnitude.}
\label{fig:mags}
\end{figure*}

	In order to define the limiting magnitude to filter the random background for the PFM fields, we made plots of the dispersion of $\theta$ in
function of the maximum $m_V$ considered. We varied the limiting magnitude with an increment of $0.1$~mag to construct the plots. The limiting
magnitude was defined as the value where the curve linearly starts to increase, after having a saturation value (SMC12 and SMC20), or abruptly starts to increase (SMC13 and SMC23). Figure \ref{fig:mags}
shows these plots for the PFM fields.

\begin{table*}[!htb]\centering
\caption{$F$-test}
\begin{threeparttable}
\begin{tabular}{ccccccc}\toprule\toprule

SMC & Model 1 & Model 2 & $F_{\text{data}}$\tnote{a} & dof\tnote{b} & $F_{\text{crit}}$\tnote{c} & Appropriate? \\\midrule

01 & 2 Gaussians & 3 Gaussians &  0.49 & (3,9) & 5.08 & yes \\ 
02 & 2 Gaussians & 3 Gaussians & -1.17 & (3,9) & 5.08 & yes \\
03 & 1 Gaussian  & 2 Gaussians & -3.93 & (3,12) & 4.47 & yes \\ 
04 & 2 Gaussians & 3 Gaussians & -1.34 & (3,9) & 5.08 & yes \\
05 & Uniform     & 1 Gaussian  & -1.85 & (2,14) & 4.86 & yes \\
06 & 1 Gaussian  & 2 Gaussians & -3.96 & (3,12) & 4.47 & yes \\
07 & 2 Gaussians & 3 Gaussians & -2.71 & (3,9) & 5.08 & yes \\
08 & 2 Gaussians & 3 Gaussians & -3.00 & (3,9) & 5.08 & yes \\
09 & 1 Gaussian  & 2 Gaussians & -1.77 & (3,12) & 4.47 & yes \\
10 & 2 Gaussians & 3 Gaussians & -3.00 & (3,9) & 5.08 & yes \\
11 & 1 Gaussian  & 2 Gaussians & -1.21 & (3,12) & 4.47 & yes \\
12 & 1 Gaussian  & 2 Gaussians & -2.74 & (3,9) & 5.08 & yes \\
13 & 1 Gaussian  & 2 Gaussians & -2.16 & (3,11) & 4.63 & yes \\
14 & Uniform     & 1 Gaussian  & -7.50 & (2,15) & 4.76 & yes \\
15 & 1 Gaussian  & 2 Gaussians & -1.88 & (3,12) & 4.47 & yes \\
16 & 2 Gaussians & 3 Gaussians & -0.87 & (3,6) & 6.60 & yes \\
17 & 2 Gaussians & 3 Gaussians & -1.88 & (3,6) & 6.60 & yes \\
18 & 1 Gaussian  & 2 Gaussians & 3.00 & (3,12) & 4.47 & yes \\
19 & Uniform     & 1 Gaussian  & -0.07 & (2,14) & 4.86 & yes \\
20 & 1 Gaussian  & 2 Gaussians & -1.58 & (3,10) & 4.83 & yes \\
21 & 1 Gaussian  & 2 Gaussians & -0.22 & (3,9) & 5.08 & yes \\
22 & 2 Gaussians & 3 Gaussians & -1.99 & (3,8) & 5.42 & yes \\
23 & 1 Gaussian  & 2 Gaussians & -1.98 & (3,7) & 5.89 & yes \\
24 & 1 Gaussian  & 2 Gaussians & -2.65 & (3,8) & 5.42 & yes \\
25 & Uniform     & 1 Gaussian  & -6.50 & (2,13) & 4.96 & yes \\
26 & 1 Gaussian  & 2 Gaussians & -2.33 & (3,7) & 5.89 & yes \\
27 & 2 Gaussians & 3 Gaussians & -2.33 & (3,7) & 5.89 & yes \\
28 & 1 Gaussian  & 2 Gaussians & -1.26 & (3,12) & 4.47 & yes \\\bottomrule

\end{tabular}

\begin{tablenotes}
{\footnotesize
\item[a] $F$ value obtained from the data.
\item[b] Degrees of freedom ($p_2-p_1$, $n-p_2$). 
\item[c] $F$ critical value.
}
\end{tablenotes}

\label{tab:ftest}
\end{threeparttable}
\end{table*}

~
~

\end{document}